\documentclass[10pt]{iopart}

\usepackage{amssymb}
\usepackage{dcolumn} 
\usepackage{graphicx}
\usepackage{hyperref}
\usepackage{color}
\usepackage{colortbl}
\usepackage[tabel]{xcolor}
\usepackage{braket}
\usepackage{multirow}
\usepackage{subfigure}
\usepackage{bm}
\usepackage{bbold}

\newcommand{\Br}{\mathbf{r}}

\newcommand{\bK}{\mathbb{K}}

\newcommand{\rev}[1]{{\color{black}{{#1}}}}

\renewcommand{\t}[1]{\textrm{\scriptsize #1}}

\newcommand{\bw}{\mathbf{w}}

\graphicspath{
  {figures/}
}

\begin{document}
\title[Gaussian Process Regression for Raman Spectra]{Using Gaussian Process Regression to Simulate the Vibrational Raman Spectra of Molecular Crystals}
\author{Nathaniel Raimbault$^{1}$\footnote{These authors contributed equally to the manuscript.}, Andrea Grisafi$^{2}\ddagger$, Michele Ceriotti$^2$, Mariana Rossi$^1$}
\address{$^1$ Fritz Haber Institute of the Max Planck Society, Faradayweg 4-6, 14195 Berlin, Germany}
\address{$^2$ \'Ecole Polytechnique F\'ed\'erale de Lausanne, Route Cantonale, 1015 Lausanne, Switzerland}
\ead{michele.ceriotti@epfl.ch and rossi@fhi-berlin.mpg.de}

\begin{abstract}
Vibrational properties of molecular crystals are constantly used as structural fingerprints, in order to identify both the chemical nature and the structural arrangement of molecules. The simulation of these properties is typically very costly, especially when dealing with response properties of materials to e.g. electric fields, which require a good description of the perturbed electronic density.
In this work, we use Gaussian process regression (GPR) to predict the static polarizability and dielectric susceptibility of molecules and molecular crystals. We combine this framework with ab initio molecular dynamics to predict their anharmonic vibrational Raman spectra. We stress the importance of data representation, symmetry, and locality, by comparing the performance of different flavors of GPR. In particular, we show the advantages of using a recently developed symmetry-adapted version of GPR. As an examplary application, we choose Paracetamol as an isolated molecule and in different crystal forms. We obtain accurate vibrational Raman spectra in all cases with fewer than 1000 training points, and obtain improvements when using a GPR trained on the molecular monomer as a baseline for the crystal GPR models.
Finally, we show that our methodology is transferable across polymorphic forms: we can train the model on data for one structure, and still be able to accurately predict the spectrum for a second polymorph. This procedure provides an independent route to access electronic structure properties when performing force-evaluations on empirical force-fields or machine-learned potential energy surfaces.
\end{abstract}
%
%
\submitto{\NJP}
%
%
\ioptwocol

\section{Introduction}
Machine-learning (ML) models are becoming increasingly popular in the field of atomistic simulations, providing a way to 
obtain data-driven physical insights~\cite{AlberiMaterials2019,GhiringhelliPRL2015,ceri19jcp} and reduce the cost of simulations~\cite{ButlerReviewML2018,BehlerReviewJCP2016}. 
Most efforts have been concentrated into predicting total energies and forces from atomic coordinates~\cite{behler2007,bartok2010,jain2013,calderon2015,BehlerReviewJCP2016,ward2017,Li,Glielmo2017,Glielmo2018}, which are most often the largest cost in a first-principles simulation. More recently, machine-learning models have been also applied to the prediction of response properties of molecules~\cite{Bereau2015,liang2017,grisafi2018,Wilkins2019,Christensen2019}. When dealing with the response of a material to an applied field, the cost of a first-principles calculation is often larger than that of force evaluation. This is thus an area where one can also take advantage of supervised learning techniques in order to reduce the cost \textit{ab initio} simulations that make use of such response properties.

Vibrational Raman spectra are a good example of properties that requires the knowledge of the response of the system to electric field perturbations. The Raman signal is very useful to monitor phase transitions, as well as for the identification of global and local structural patterns~\cite{Agrawal_Nature_2016,Heiner_biophotonics_2018,Monserrat_PRL_2018}. Any technique used to calculate this property requires the calculation of several instances of the polarizability tensor (in molecules) or the dielectric susceptibility (in crystals). Previously, some of the present authors have shown that anharmonic vibrational Raman spectra calculated through a time-correlation formalism can be a powerful tool to identify structural fingerprints in molecular crystals~\cite{DFPT_dielectric_2018,Raimbault_PRM_2019}. Within this formalism, it is necessary to calculate \textit{ab initio} molecular dynamics trajectories and compute the response quantities for subsequent atomic configuration, employing, for instance, density-functional perturbation theory (DFPT) \cite{Gerratt_JCP_1968, Gonze_APS_1997-2, Putrino:2000fm, BaroniRMP2001, DFPT_dielectric_2018}. These calculations are computationally demanding, not only because of the tens of thousands of force evaluations that need to be performed to provide sufficient statistical sampling, but also because each DFPT calculation is typically four times more expensive than a force evaluation \cite{Raimbault_PRM_2019}\footnote{This number is clearly system- and settings-dependent, and simply represents the estimate reported in Ref.\cite{Raimbault_PRM_2019}.}. Furthermore, while there are several empirical potentials available that can be used to simulate the dynamics of molecular crystals~\cite{Reilly:2016cj}, empirical models of the polarizability tensors are rare and poorly transferable \cite{Lemkul:2016kv}.

%
\begin{figure}[tbph]
 \includegraphics[width=1.0\columnwidth]{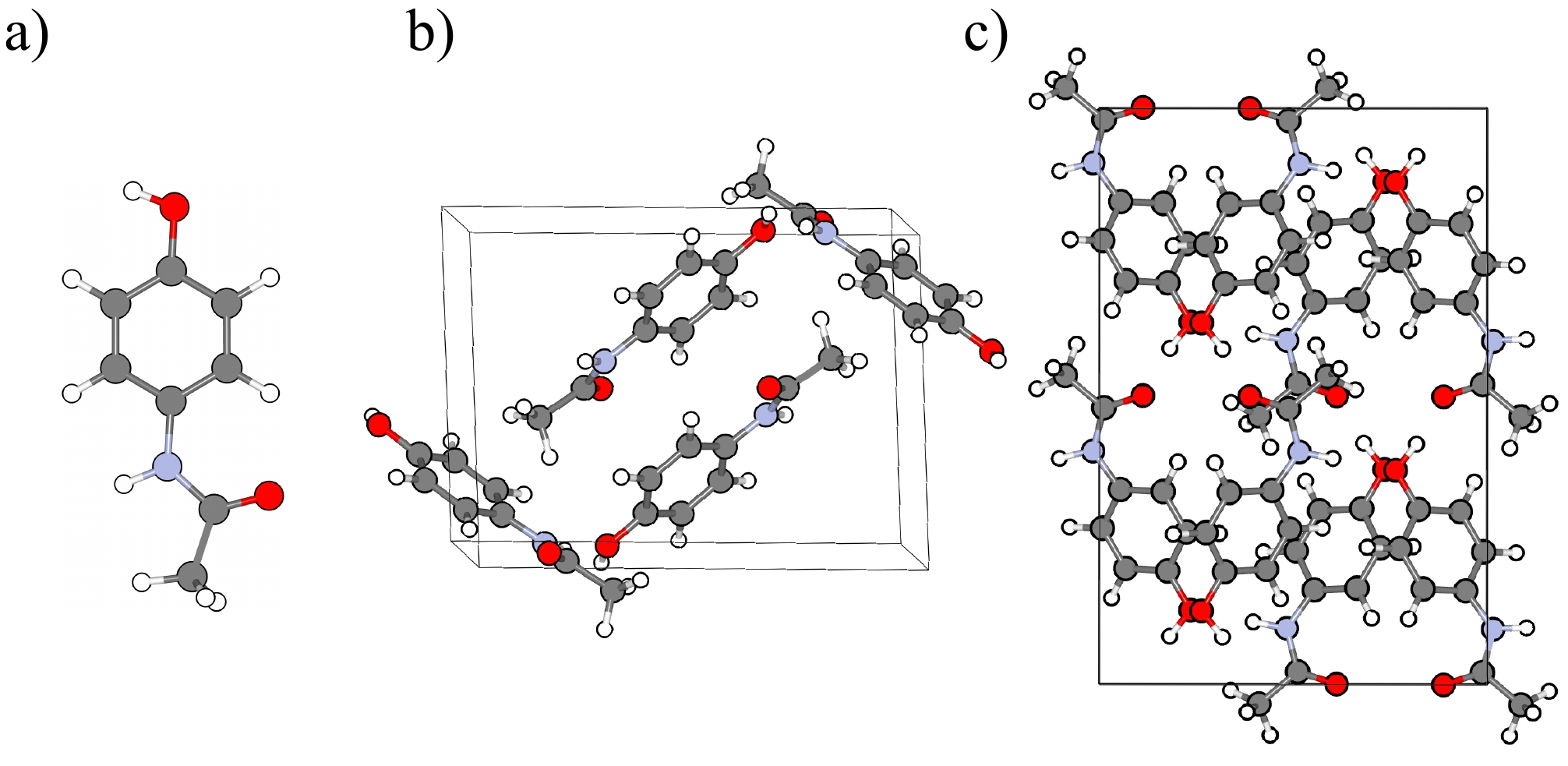}
 \caption{The systems considered in this work for the prediction of vibrational Raman spectra. (a) Isolated Paracetamol molecule. (b) Paracetamol crystal form I (monoclinic). (c) Paracetamol crystal form II (orthorhombic). Atomic color code: hydrogen white, nitrogen blue, carbon grey, oxygen red. The unit cell is drawn in black.}
 \label{fig_paracetamol}
\end{figure}

Here we investigate frameworks to obtain accurate predictions of the dielectric response properties for a number of consecutive molecular dynamics (MD) configurations, that are necessary to converge simulated vibrational Raman spectra of molecules and molecular crystals. We compare different flavors of Gaussian Process Regression (GPR), which is a method that has already been proven to be efficient in predicting dielectric response properties~\cite{liang2017,Wilkins2019,Bereau2015}. In particular, we compare standard GPR schemes with symmetry-adapted GPR (SA-GPR)\cite{Grisafi_SAGPR}, which is advantageous when describing tensorial quantities. 
For the former, we exploit the internal structural rigidity of the system in order to model each individual component of the polarizability tensor, therefore remapping the tensor learning problem onto many separate scalar regression tasks. For the latter, we employ a symmetry-adapted representation of the system in order to learn the irreducible spherical components of the tensor in a covariant fashion. 
As a trial system we consider the Paracetamol molecule and form I and form II of the Paracetamol crystal, represented in Fig.~\ref{fig_paracetamol}. As we demonstrate below, SA-GPR comes out as the methodology with the best performance. When predicting the Raman spectra of crystals, both methods can benefit from using a GPR trained on the monomer as a baseline. We also find that our model is transferable between different polymorphic forms.  Given that empirical and machine-learned potential energy surfaces are becoming more accurate for molecular crystals, the methodology proposed here can be combined in a straightforward manner to such potentials, giving access to the electronic polarization and polarizability of crystals.

In the following, Section \ref{sec:theory} introduces the general machinery describing GPR and SA-GPR, as well as the different representations used to map structures to the inputs of the ML scheme.
In Section \ref{sec:results}, we present applications on the Paracetamol molecule and monoclinic form I crystal, showing for the latter how our models can be refined by the inclusion of molecular polarizability tensors.
Finally, we illustrate in Section \ref{sec:transfer} the transferability of the SA-GPR method, by predicting the Raman spectrum of the orthorhombic form II crystal without prior knowledge of the corresponding values of the dielectric response.

\section{Theory}

\label{sec:theory}

\subsection{Vibrational Raman Spectra}

The central quantity needed for simulating vibrational Raman spectra is the static polarizability tensor $\bm{\alpha}$. For simplicity, we will refer to the polarizability tensor for all the rest of the paper, but one should keep in mind that for solids the quantity of interest is rather the electric susceptibility tensor of the system. 

As discussed in Ref.~\cite{Raimbault_PRM_2019} and others \cite{Putrino_Parrinello_PRL_2002,Pagliai_JCP_2008}, the vibrational Raman spectrum can be calculated using several approximations, the simplest of which is the harmonic approximation. The framework developed in this paper can be applied to the harmonic case, as shown in the supplemental information, Fig. S8, but we will here focus on the more challenging task of applying it to the linear-response time-correlation formalism, which fully takes into account the anharmonicity of the potential energy surface.
In this formalism, vibrational Raman intensities can be obtained from the Fourier transform of the static polarizability autocorrelation function~\cite{Mukamel_1995} at thermodynamic equilibrium. 
In particular, the so-called powder spectrum intensity is given by a combination of the anisotropic and isotropic contributions as %
\begin{eqnarray} 
 I(\omega) &= & I_\t{iso}(\omega) +\frac73 I_\t{aniso} (\omega) \\
\nonumber I_\t{iso}(\omega) & = &\frac{n}{2\pi} \int_{-\infty}^{+\infty}dt e^{-i\omega t} \langle\bar{\alpha}(0)\bar{\alpha}(t)\rangle  \\
\nonumber I_\t{aniso} (\omega) & = &\frac{n}{2\pi} \int_{-\infty}^{+\infty}dt e^{-i\omega t} \frac{1}{10}\langle \textrm{Tr}[\bm{\tilde{\alpha}}(0)\cdot \bm{\tilde{\alpha}}(t)]\rangle \,,\label{eq_raman_intensity}
\end{eqnarray}
where $n$ is the number of atoms in the system, the brackets $\langle\cdot\rangle$ denote an ensemble average and $\textrm{Tr}$ represents the trace. $\bar{\alpha}$ and $\bm{\tilde{\alpha}}$ are, respectively, the isotropic and anisotropic parts of the polarizability tensor, defined as follows,
\begin{eqnarray}
\nonumber \bm{\alpha}&=&\bar{\alpha} \bm{I}+\bm{\tilde{\alpha}} \\
 \bar{\alpha}&=&\frac13(\alpha_{xx}+\alpha_{yy}+\alpha_{zz}) ~,~
 \textrm{Tr}[\bm{\tilde{\alpha}}]=0 ~.
 \label{eq_alpha_decomposition}
\end{eqnarray}

\rev{In this paper, we do not address the problem of obtaining forces, which are necessary to calculate a full Raman spectrum from either molecular dynamics trajectories or in the harmonic approximation. Instead, we focus on predicting only the Raman intensities and combine them with pre-computed \textit{ab initio} trajectories and displacements.}

\subsection{Component-wise Gaussian Process Regression}
\label{subsec:GPR}

Gaussian process regression (GPR) is a well-established method based on a kernel function that measures the similarity between structures.
\rev{It is formally equivalent to kernel-ridge regression (KRR), to which it differs mostly by the fact that GPR frames the construction of the model in a Bayesian language.}
In this sense, the kernel represents the prior distribution for the statistical correlations of the property we aim to predict. In the usual supervised-learning framework, a dataset of structures (i.e., atomic coordinates) and associated polarizabilities is used to train the model. Once the training is complete, the model is tested on a different set of configurations for which the polarizability is in principle not known. If it is possible to align the system to a reference structure, a straightforward way to apply this procedure to tensorial quantities is to learn each component of the polarizability tensor separately. In particular, a GPR prediction of each individual polarizability component $\alpha_{\gamma \delta}$ reads
\begin{equation}
 \alpha^\t{ML}_{\gamma \delta}(\mathcal A)=\bar{\alpha}^\t{ai}_{\gamma \delta}+\sum_{j=1}^{N} w_{j}^{\gamma\delta} k(\mathcal A,\mathcal A_j) \,,
 \label{eq_GPR}
\end{equation}
where $N$ is the number of configurations included in the training set, $\gamma$ and $\delta$ represent Cartesian coordinates, $w_{j}^{\gamma\delta}$ are the regression weights that need to be determined from the training data for each component, and $k$ is the kernel that couples the target system $\mathcal A$, with the training structure $\mathcal A_j$.  The quantity $\bar{\alpha}^\t{ai}_{\gamma \delta}$ is the average (over the training set) of the $\gamma \delta$ polarizability component computed from an \textit{ab initio} method, which effectively allows the training to focus on the fluctuation of the property with respect to a known baseline value.

The kernel entering GPR is based on a Gaussian similarity measure between structures $\mathcal A$, given by
\begin{equation} 
 k(\mathcal A,\mathcal A_j)= e^{-\frac{|\boldsymbol{u}(\mathcal A)-\boldsymbol{u}(\mathcal A_j)|^2}{2\sigma^2}} ,\label{eq:gkernel}
\end{equation}
with $\sigma$ being a hyperparameter that controls the magnitude of the correlation between training points. 
\rev{We optimize the value of $\sigma$ - as well as that of other hyperparameters entering the definition of the features or the kernel - by cross-validation (CV), even though in a GPR framework hyperparameters are often determined by likelihood maximization (see SI, section~1 for further details).}
$\boldsymbol{u}$ is a vector that has the role of mapping the atomic coordinates of the structure $\mathcal A$ to a given representation of dimension $M$. 

The regression weights $\bw^{\gamma\delta}$ are obtained by minimizing a loss function regularized by an $L2$-norm~\cite{Vu_IJQC_2015} over the training set. This procedure leads to the following expression
\begin{equation}
 \bw^{\gamma\delta}=\left(\bK+\eta \mathbb{1} \right)^{-1}\cdot \Delta\boldsymbol{\alpha}_{\gamma \delta} \,,
\end{equation}
where $\mathbb{1}$ is the identity matrix, $\mathbb{K}$ is the $N \times N$ kernel matrix associated with the reference structures, such that $\mathbb{K}_{ij}=k(\mathcal A_i,\mathcal A_j), i,j=1,... ,N$, and $\eta$ is the regularization parameter which controls to which extent the fitted data can deviate from the training points.
The quantity $\Delta\boldsymbol{\alpha}_{\gamma \delta}$ represents the vector containing all $N$ entries in the training set of $\Delta \alpha_{\gamma \delta}^j \rev{(\mathcal A_j)}=\alpha^\t{ai}_{\gamma \delta}(\mathcal A_j)-\bar{\alpha}^\t{ai}_{\gamma \delta}$.

The efficiency of any GPR model strongly depends on the quality of the representation that is encoded in $\boldsymbol{u}(\mathcal A)$ (see Eq. \ref{eq_GPR}). For a representation to be efficient, it should contain the least possible number of elements to express the identity of a given structure, while avoiding redundant information.
\rev{Among the vast choice of representations one could conceive, we will employ two that are very similar in spirit. First, we will consider a representation of each structure in terms of its atomic density (AD), explicitly evaluated on a grid. This AD is combined with the GPR framework, and with an alignment procedure that makes it possible to learn tensor components independently.  Second, we will consider a symmetry-adapted Gaussian process regression (SA-GPR) scheme~\cite{Grisafi_SAGPR},   whose application is summarized in the next section. The framework is based on $\lambda$-SOAP kernels, that are built by covariant integration of the atom density~\cite{will+19jcp}, and therefore automatically incorporate molecular symmetries.}

\begin{figure}[htbp]
  \centering
  \includegraphics[width=1.0\columnwidth]{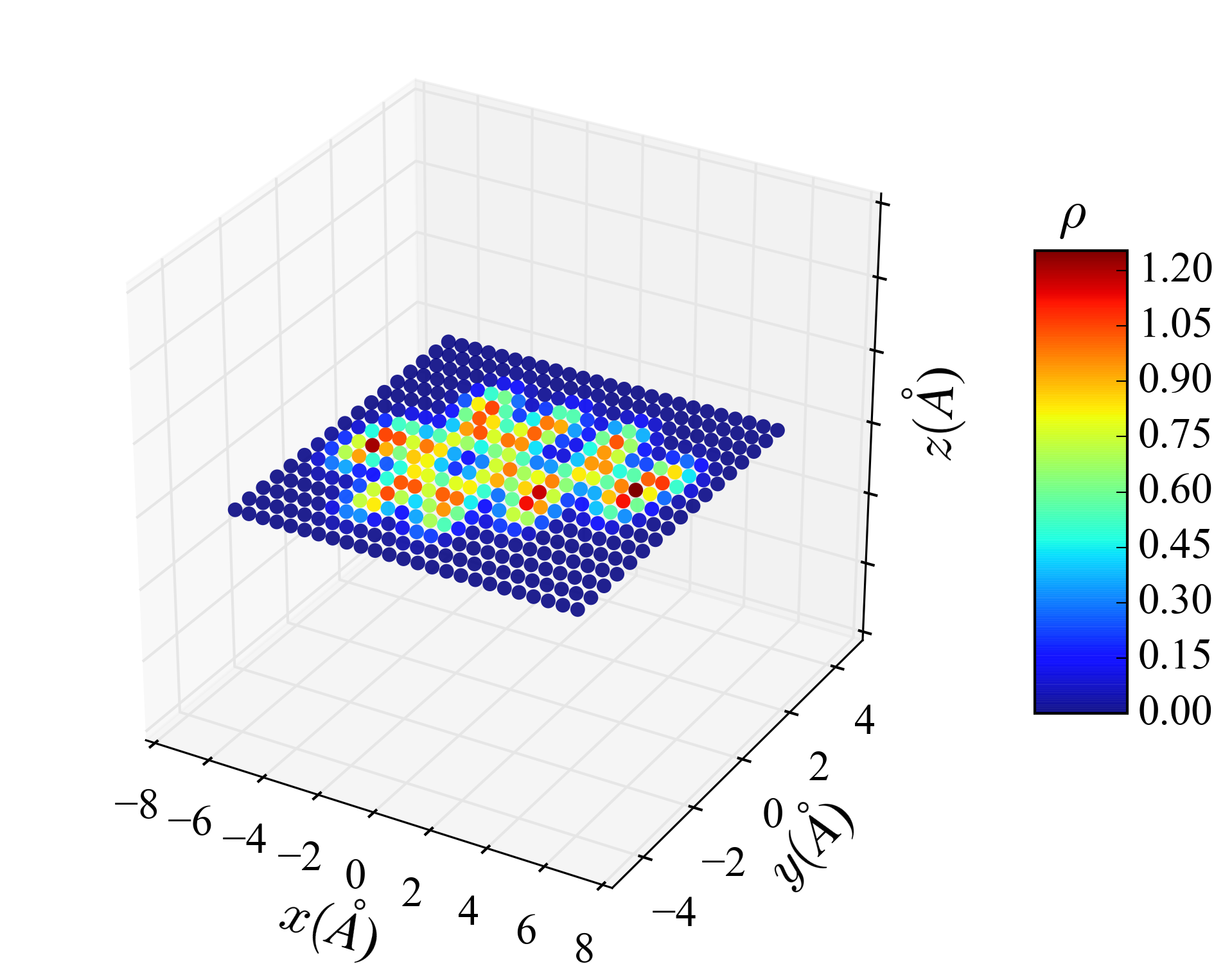}
  \caption{2D view of the nuclear density distribution of the Paracetamol molecule for a given structure. Each grid point is represented by a colored sphere. Blue (red) indicates a low (high) density.}
  \label{fig_den_dist}
\end{figure}

Atomic-density grids consist of a conceptually simple representation. 
The construction introduces a Cartesian reference frame centered on the system under study, and defines a 3-dimensional grid around it. For each grid-point $\Br$, we calculate the atomic density distribution, defined as
\begin{equation} \label{eq_AD_species}
 \rho_s(\Br) =\sum_{i \in s} \exp\left(-\frac{|\Br-\Br_i|^2}{2\gamma_{s}^2}\right) \,,
\end{equation}
where $s$ identifies a specific atom type, $\Br_i$ are the nuclear positions and $\gamma_{s}$ is an adjustable parameter. The feature vector is given by $\boldsymbol{u}(\mathcal A) = (\rho_s(\Br)), s=1,...,N_s$, where $N_s$ is the number of different atomic species.  For the applications \rev{using the standard GPR scheme} shown in this paper, we have chosen the same $\gamma_{s}=0.5$ {\AA} for all species. \rev{This choice was based on the fact that we observed very minute changes when employing different values of $\gamma_s$ for different species, but this observation is likely to be system- and method- dependent}. 
We have used a grid of evenly-spaced points spanning the maximal extension of the system, although we note that more refined methods could be utilized to define physically-motivated grid points, based on the possible directions of vibrations of atomic species~\cite{Zauleck_JCTC_2018}.
Such a representation is illustrated in Fig.~\ref{fig_den_dist}, which shows a 2-dimensional cut of the density distribution associated with a paracetamol molecule. 
\rev{Note that one could use a combination of different descriptors instead of a single one, whose elements would be concatenated to form the feature vector $\boldsymbol{u}$.}

As mentioned above, the components of the polarizability tensor are not invariant to rotations in cartesian space, and therefore a Cartesian space representation of this quantity, like the atomic densities, requires an alignment to a reference structure.
To do so, we have used the Kabsch algorithm considering only heavy atoms~\cite{Kabsch_ACSA_1976,Kabsch_ACSA_1978}.
We note that this alignment procedure is not applicable to all systems, in particular to very flexible molecules. For simple relatively rigid molecules, it is known to work well, as has been shown previously for the case of hyperpolarizabilities of water molecules~\cite{liang2017}. 
We will show below that even for a more complex and relatively flexible molecule like paracetamol, this representation still yields accurate predictions.

\subsection{Symmetry-adapted Gaussian Process Regression with $\lambda$-SOAP Kernels}

SA-GPR~\cite{Grisafi_SAGPR} represents a generalization of the GPR formalism, where the tensorial nature of the target property, together with its covariant 3D transformations, are naturally incorporated within the regression algorithm~\cite{Glielmo2017}. This technique removes the need of often arbitrary alignment procedures of the systems (like the one described above) in order to predict tensorial quantities of any rank.  
As such, the model focuses only on the portion of variability of the tensor connected with an internal structural distortion of the molecular geometry, greatly improving the regression performances. 
\rev{Furthermore, by effectively learning an atom-centered model of the polarizability, a SA-GPR scheme makes it possible to build models that are transferable between different molecules~\cite{Wilkins2019}, although we didn't exploit this possibility in the present work.}

A simplification of the learning problem can be obtained if the target property is first decomposed in its irreducible spherical tensor components. The static polarizability (or the static susceptibility), in particular, being a symmetric rank-2 tensor, can be formally decomposed into an isotropic contribution that transforms as a spherical harmonic of angular momentum $\lambda=0$, and an anisotropic contribution that transforms as a spherical harmonic of angular momentum $\lambda=2$. The former contribution, $\alpha^{(0)}_{0}\propto\bar{\alpha}$, being directly proportional to the trace of the tensor (which is rotationally and translationally invariant), can be learned in the usual manner by a standard GPR. 
The tensorial nature of the polarizability is contained in the $\lambda=2$ term,  $\boldsymbol{\alpha}^{(2)}=(\alpha^{(2)}_{-2},\alpha^{(2)}_{-1},\alpha^{(2)}_{0},\alpha^{(2)}_{1},\alpha^{(2)}_{2})$, which is related to the anisotropic part of the Cartesian $\boldsymbol{\alpha}$-tensor $\bm{\tilde{\alpha}}$ of  Eq.~\ref{eq_alpha_decomposition} by a linear transformation.
Within SA-GPR, the prediction of this contribution is carried out by making use of a tensorial kernel function $\boldsymbol{k}^{(2)}(\mathcal A,\mathcal B)$, which is a matrix of size 5$\times$5, that describes at the same time the structural similarity between molecular configurations and the tensorial geometric relationship of order $\lambda=2$ between these configurations, as detailed in Ref.~\cite{Grisafi_SAGPR}. In general terms, such a kernel can be thought of as a generalization of a scalar kernel function such as the one introduced in Eq. \ref{eq:gkernel}, which makes use of the following covariant integration,
\begin{equation}\label{l-soap}
   \boldsymbol{k}^{(2)}(\mathcal A,\mathcal B) = \int \mathrm{d}\hat{R}\ \boldsymbol{D}^{(2)}(\hat{R})\kappa(\mathcal A,\hat{R}\mathcal B)  \,,
\end{equation}
where $\hat{R}$ represents the rotation operator and $\boldsymbol{D}^{(2)}(\hat{R})$ the associated Wigner-D matrix which has the role of expressing the rotation of $\lambda=2$ spherical harmonics.
In this definition, the scalar kernel $\kappa(\mathcal A, \mathcal B)$ only needs to be invariant under rigid translations and rotations of the laboratory reference frame with respect to which both $\mathcal A$ and $\mathcal B$ are defined. In the present work, $\kappa(\mathcal A,\mathcal B)$ is given by a superposition of atom-centered Gaussian densities, equivalent to the one used in the popular Smooth Overlap of Atomic Position (SOAP) kernel~\cite{bartok2013}. As such, the kernel of Eq.~(\ref{l-soap}) represents the tensorial generalization of SOAP, usually called $\lambda$-SOAP, which recovers the scalar case of $\lambda=0$ as a special limit~\cite{grisafi2018}. 

Formally, a covariant $\lambda=2$ prediction performed with a symmetry-adapted kernel function of the same order reads
\begin{eqnarray}
  \nonumber  \boldsymbol{\alpha}^{(2)}(\mathcal{A}) &= \sum_{i=1}^N \boldsymbol{k}^{(2)}_{\mathcal A,i} \ \boldsymbol{w}^{(2)}_i \\
    &= \sum_{i=1}^N \boldsymbol{k}^{(2)}_{\mathcal A,i} \sum_{j=1}^N \left[\bK^{(2)}+\eta\mathbb{1}\right]^{-1}_{ij}\boldsymbol{\alpha}^{(2)}_j \,,
\end{eqnarray}
where $i$ and $j$ run over the $N$ reference configurations used to train the regression model and $\eta$ is the regularization parameter. The set of tensorial regression weights $\{\boldsymbol{w}^{(2)}_i\}$ are determined by inverting the $(5\times N)^2$ kernel matrix $\bK^{(2)}$ associated with the training structures, and projecting it on the vector of reference tensors
$\{\boldsymbol{\alpha}^{(2)}_j\}$. In doing so, the statistical average of $\{\boldsymbol{\alpha}^{(2)}_j\}$ over the training set is assumed to be zero by symmetry. As such, no baseline of the anisotropic part of the polarizability is adopted when doing the regression.

\subsection{Errors and uncertainty estimations \label{subsec:errors}}

In order to gauge the accuracy of the machine-learned polarizability components, we calculate the root mean square error (RMSE) normalized by the standard deviation (STD) of the set we want to evaluate the error on,
\rev{
\begin{eqnarray}
\epsilon_{\gamma \delta}[\sigma,\eta]= 100\times \sqrt{ \frac{\frac{1}{N}\sum_{j}^N \left(\alpha_{\gamma \delta}^{\t{ML}}(\mathcal A_j)-\alpha_{\gamma \delta}^{\t{ai}}(\mathcal A_j)\right)^2}{\frac{1}{N-1}\sum_{j}^N (\alpha^\t{ai}_{\gamma \delta}(\mathcal A_j)-\bar{\alpha}^\t{ai}_{\gamma \delta})^2}} \,, 
\label{eq_error}
\end{eqnarray}}
where $j$ represents the $j$th configuration, $N$ is the amount of points in the data set we consider, and the bar here denotes the average of the polarizability component over the dataset of interest. 

Whenever the reference properties of the testing data are not available, one cannot make use of Eq.~(\ref{eq_error}) to evaluate the error of the predicted polarizabilities. In these circumstances, one typically needs to estimate the error made on the predicted properties by making use of some \emph{a priori} probabilistic criteria. In the particular case of GPR, the expected error associated with a testing structure $\mathcal A$, can be computed as $\varepsilon(\mathcal A) = k(\mathcal A,\mathcal A)-\sum_{IJ}k(\mathcal A,I)k^{-1}(I,J)k(J,\mathcal A)$, where $I$ and $J$ run over the training structures. As detailed in Ref.~\cite{Musil_JCTC_2019}, this strategy is however not very practical because of its computational expense, so that other kind of methods such as bootstrapping or subsampling can rather be used to estimate the prediction errors. In addition, in the particular case of the present work one would like to propagate the error that occurs in the prediction of $\boldsymbol{\alpha}$ to the Raman spectrum. While this error propagation would be difficult to carry out on top of the GPR intrinsic covariance, it comes naturally from a bootstrapping or subsampling scheme. 

In this work, $N_{\t{RS}}$ subselections of the training dataset are considered to generate an ensemble of predictions for the polarizability. From these,  $N_{\t{RS}}$ Raman spectra are computed by Fourier transforming the time series of each model in the ensemble. Finally, the average and the standard deviation of the predicted spectra over the $m$ subselections give the final Raman spectrum prediction and the propagated estimated error respectively.  
The downside of this approach is that this model works under the assumption that the training data corresponds to independent identically distributed samples. This is of course not true in general, so that one needs to correct the model to take into account for the underlying correlations. 
Following Ref.~\cite{Musil_JCTC_2019}, a maximum likelihood recipe can be adopted to linearly scale the variance of the predictions by a constant factor $\nu^2$. The calibration of this scaling factor is carried out by computing the actual prediction errors of the polarizabilites over a suitably selected validation set $N_\t{val}$, for which the reference polarizabilities are known, and then considering  
\begin{equation}
    \nu^2 = \frac{1}{N_{\t{val}}}\sum_{j=1}^{N_{\t{val}}} \frac{||\boldsymbol{{\alpha}}_\t{pred}(j)-\boldsymbol{{\alpha}}_\t{ref}(j)||^2}{\sigma^2(j)}
\end{equation}
where $\sigma^2(j)$ are the variances of the predicted polarizabilities.
Once the value of $\nu$ has been determined, each polarizability prediction of a given training model $k$ can be updated by considering
\begin{equation}
    \boldsymbol{\alpha}'_k = \bar{\boldsymbol{\alpha}} + \nu(\boldsymbol{\alpha}_k-\bar{\boldsymbol{\alpha}})
\end{equation}
where $\bar{\boldsymbol{\alpha}}$ is the predicted polarizability averaged over the $N_{\t{RS}}$ models. This scaling procedure guarantees that the variance of the models is consistent with the outcome of the likelihood maximization. By computing the Raman spectrum for each scaled model $k$, the propagated uncertainty estimation associated with the spectra will automatically take into account the calibration of the variance.

\subsection{Simulation and training details}

To perform the \textit{ab initio} calculations, we used the FHI-aims~\cite{Blum_2009} program package with \textit{light} settings for all atomic species. We obtained aiMD trajectories using the PBE functional with many-body dispersion (MBD) corrections~\cite{Tkatchenko_PRL2012,Ambrosetti_JCP2014}, and employing a time step of 0.5 fs. The polarizabily/susceptibility was instead computed every 1 fs. Most of the data used here was already available from references \cite{DFPT_dielectric_2018} and \cite{Raimbault_PRM_2019}.
For each system, we had 20 picoseconds of simulation in the NVT ensemble at 300 K. From this trajectory, a few thousands of configurations were selected to define the training and test set of our model. A full trajectory of 15 picoseconds in the NVE ensemble was instead considered as our validation set to assess the quality of our predictions, by comparing the predicted Raman spectra to the ones obtained with \textit{ab initio} polarizabilities. We chose to train our model on the NVT ensemble and predict on the NVE ensemble. This choice is justified by the fact that - particularly for this relatively short trajectory - canonical sampling should be more ergodic than microcanonical sampling, and therefore sample a larger portion of configurational space. 

\section{Results}
\label{sec:results}

In the following, we apply the previously described methods to the calculation of the Raman spectra of Paracetamol.
Each time we mention GPR, it is used in combination with the AD representation. Similarly, each time SA-GPR is mentioned, it is associated to $\lambda$-SOAP kernels.

\subsection{Paracetamol molecule}

\begin{figure*}[htbp]
  \subfigure[]{
   \includegraphics[width=1.0\columnwidth]{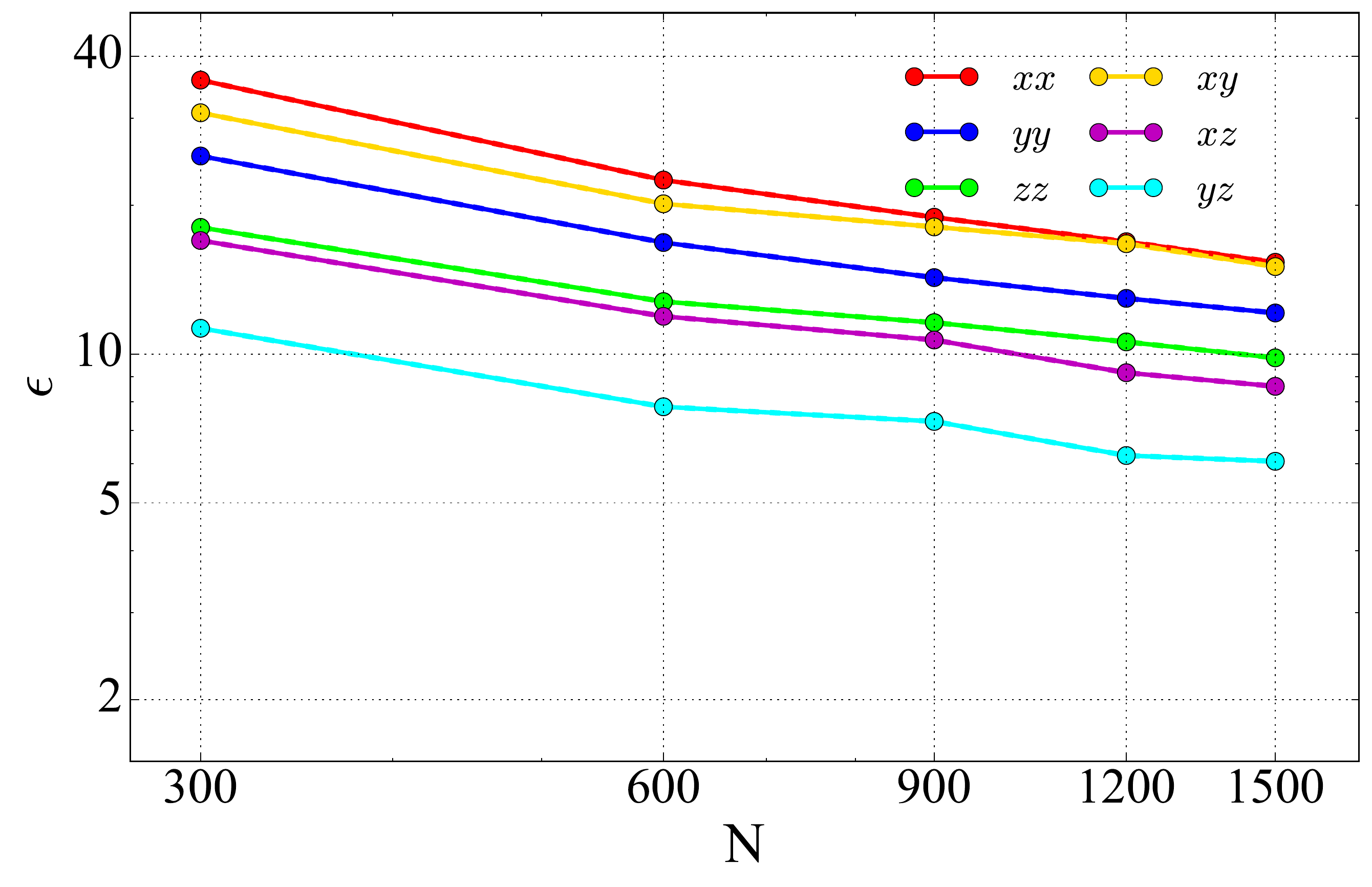}
   }
  \subfigure[]{
   \includegraphics[width=1.0\columnwidth]{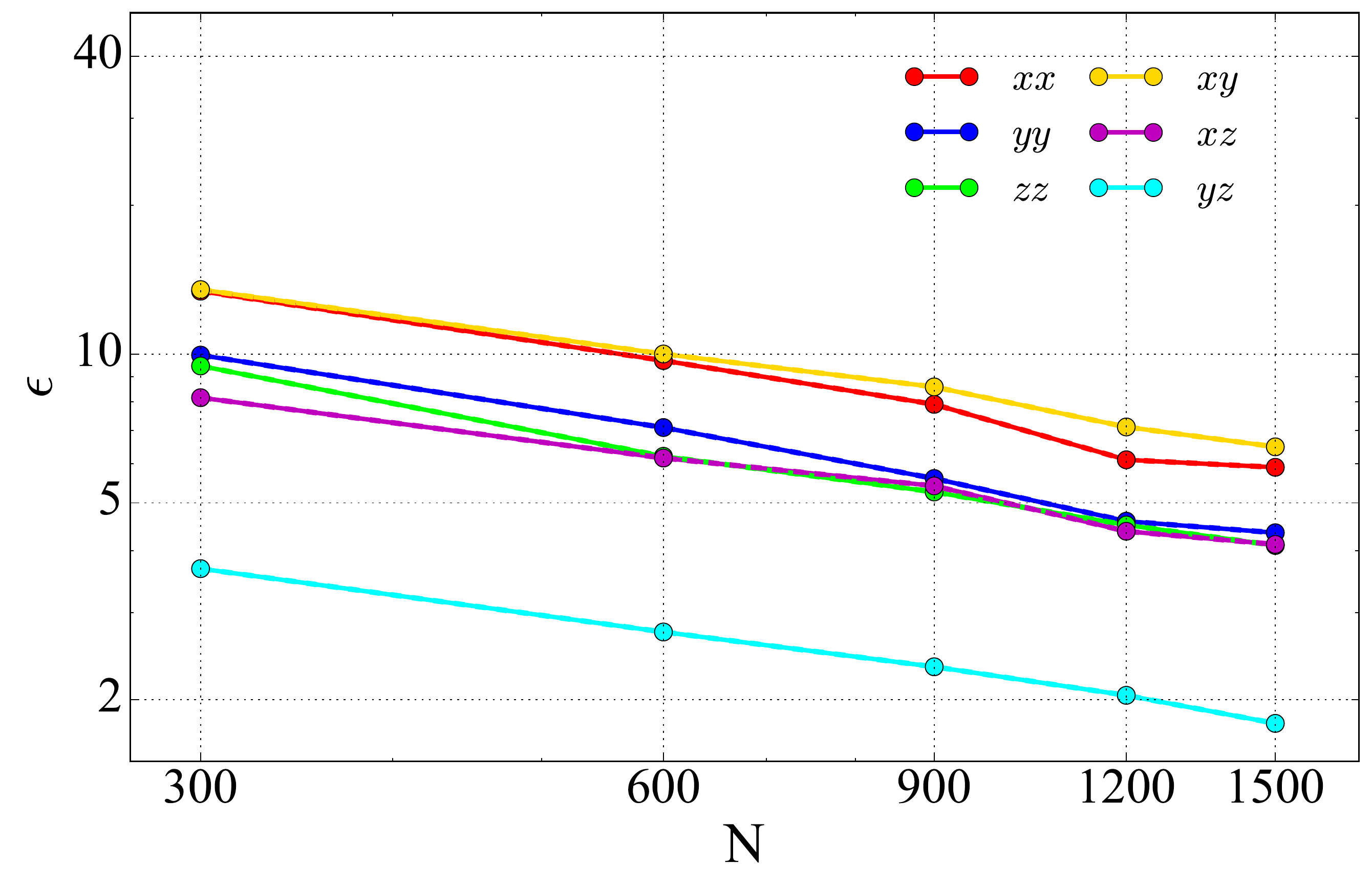}
  }
  \caption{Prediction error (as defined by Eq. \ref{eq_error}) on each component of the polarizability tensor of the paracetamol molecule with different models. (a) Learning curves from GPR using atomic densities as a representation. (b) Learning curves from SA-GPR with $\lambda$-SOAP kernels.}
 \label{fig_learningcurve_molecule}
\end{figure*}

We first consider the case of the isolated Paracetamol monomer (Fig.~\ref{fig_paracetamol} (a)). We constructed the training dataset by selecting 2000 structures with farthest point sampling (FPS) using the scalar SOAP metric from the full NVT trajectory. 
For GPR, the three-dimensional density field was constructed within a box of $6 \times 4 \times 2.5$ \AA, where the molecule had its longest axis along the $x$ direction and the equilibrium geometry lied on the $xy$ plane, and the grid spacing was $dr=0.5$ \AA.
The GPR was computed using $\sigma = 10$ and $\eta=10^{-3}$. 
Details about the optimization of the hyperparameters are given in the supplemental information.
Regression performances are reported in Fig.~\ref{fig_learningcurve_molecule} (a), where the error $\epsilon$, given by Eq.~\ref{eq_error}, computed on 500 randomly selected test structures (that are excluded from training) out of the total of 2000 is shown as a function of the number of training monomers.~\footnote{Learning performance when using only unprocessed atomic coordinates are also shown in the SI, Fig.~S2.}

Figure \ref{fig_learningcurve_molecule}(a) shows that the  learning capability of the model does not reach saturation within the training set sizes explored.
The learning of all the polarizability components follow a similar slope, but they are predicted with different accuracy because of the strong anisotropy of the dielectric response of the Paracetamol molecule. Because of the $\pi$-conjugation of the system in the molecular plane, the system is much more polarizable along the $x$-axis rather than along other directions in this particular alignment, making it harder for the learning algorithm to capture the corresponding variations across the dataset. 
The $\alpha_{xx}$ component presents the largest error, going from about 40\% with 300 training points to 17\% with 1500 training points. The best learning performance is instead obtained for the $\alpha_{yz}$ component, where the prediction error can be brought down to about 6\%.  

\begin{figure}[htbp]
  \centering
  \includegraphics[width=1.0\columnwidth]{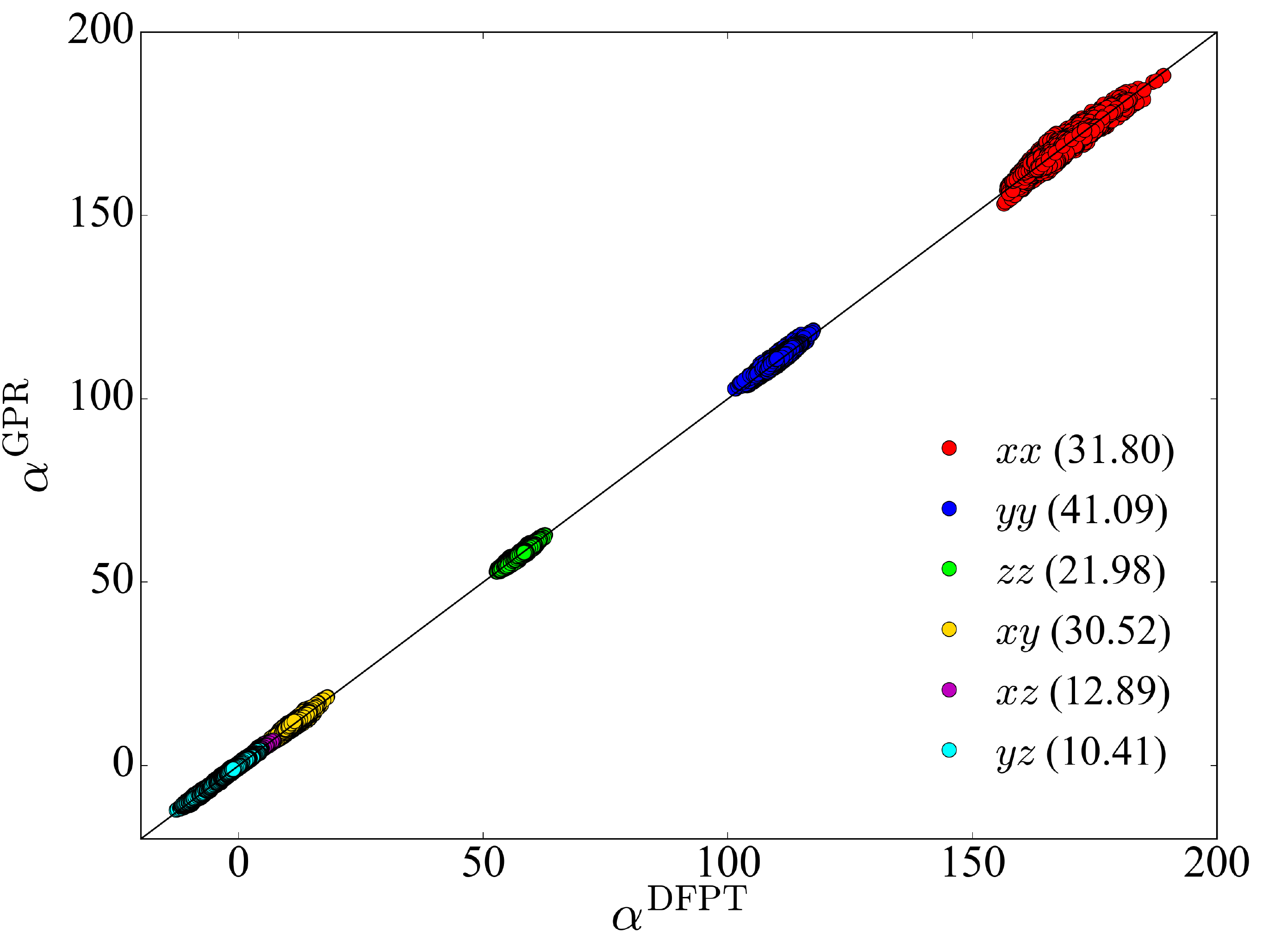}
  \caption{GPR polarizability tensor components versus DFPT \textit{ab initio} ones. The components were trained on 900 configurations coming from an NVT trajectory. The test set contains 20000 configurations coming from an NVE trajectory. Numbers in brackets indicate $\epsilon$ for a given component.}
  \label{fig_polar_abinitio_vs_GPR}
\end{figure}

In order to validate the model, we show in Fig.~\ref{fig_polar_abinitio_vs_GPR} the correlation between predicted and computed polarizability components on the validation set composed by the full independent NVE trajectory, for a representative training size of 900 molecular configurations, including the $\epsilon$ for each component. Although we observe a worsening of the predictions when compared to the previous case, where we trained and predicted on the same ensemble (and trajectory), the predicted polarizabilities are still well correlated to the reference values. The remaining question is how these errors translate to the actual prediction of the vibrational Raman spectrum.

\begin{figure*}[htb]
   \centering
   \includegraphics[width=0.9\textwidth]{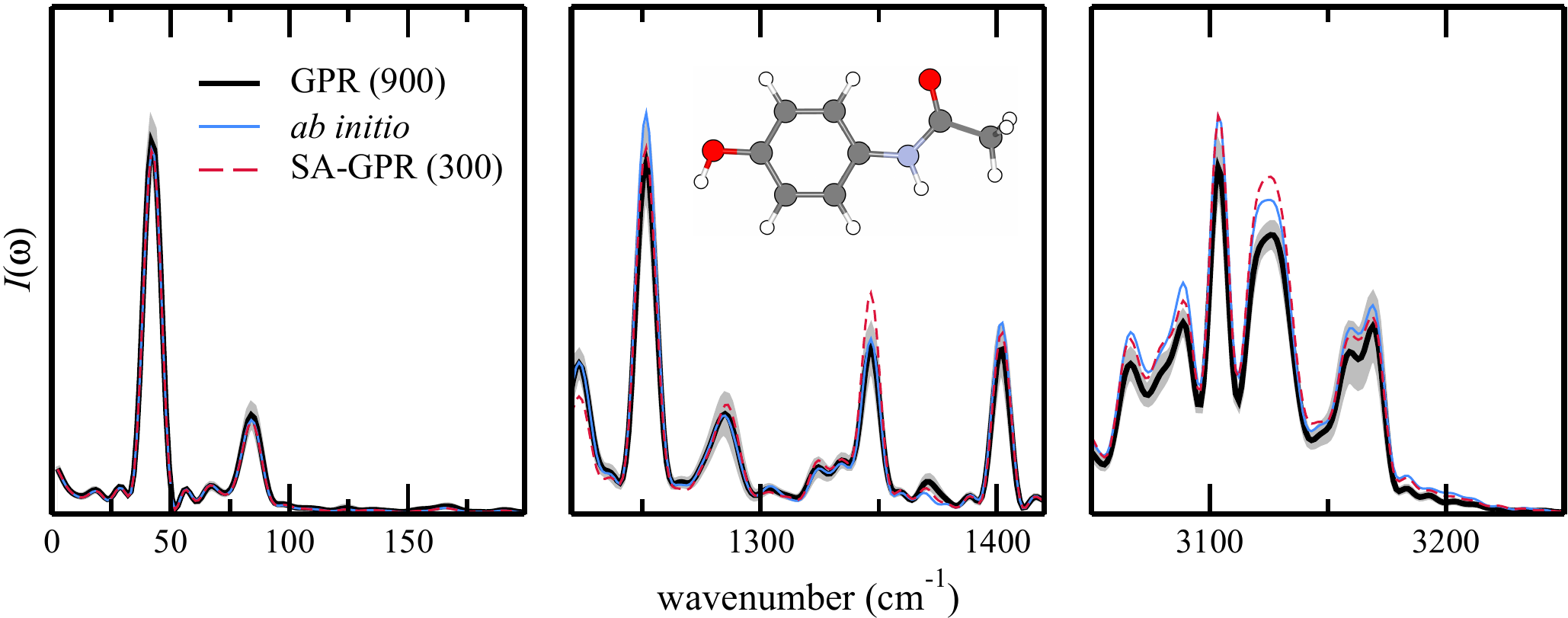}
   \caption{(\emph{black line}) Raman spectrum GPR prediction of the Paracetamol molecule averaged over 16 different training models. Each training model is obtained by a random subselection of 900 configurations over a total of 1100, while the prediction was made on 20000 structures. (\emph{shaded area}) Standard deviation of the predicted spectra over the 16 models\rev{, calibrated with a likelihood maximization procedure described in Section \ref{subsec:errors}}. (\emph{blue line}) reference \textit{ab initio} Raman spectrum. (\emph{red dotted line}) Single SA-GPR prediction using 300 training points.}
   \label{fig_Raman_GPR_molecule}
\end{figure*}

In Fig~\ref{fig_Raman_GPR_molecule}, we show a machine-learned Raman spectrum averaged over 16 subselections of the training set of 900 configurations each, along with its standard deviation (see Sec. \ref{subsec:errors} for a detailed discussion about this procedure), and compare it to the one calculated from fully \textit{ab initio} data. We find that the estimated variance has to be scaled by a factor of $\nu^2 = 2.0$. Despite a relatively small amount of training points, the agreement with the reference spectrum is excellent in the entire frequency range. As shown in Fig.~\ref{fig_learningcurve_molecule}, increasing the number of training points in the model would decrease errors even further.
From Figs. \ref{fig_learningcurve_molecule} and \ref{fig_Raman_GPR_molecule}, it is clear that the error we make on the polarizability components does not translate directly into an error of similar magnitude on the spectrum. This is a consequence of the fact that the Raman intensities depend on the derivatives of the polarizability components with respect to atomic coordinates, and not on their absolute value. 
This simple procedure is able to reproduce almost perfectly a reference Raman spectrum with fewer than 1000 training points on a desktop computer in just a few minutes.

In order to assess whether the use of a model that incorporates symmetries can be advantageous even for a relatively rigid molecular system, we then contrast the GPR and the SA-GPR model. To this end, $\lambda$-SOAP  kernels were constructed using a Gaussian width of 0.3 {\AA} and an environment cutoff of 4.0 {\AA}. \rev{Details about the SOAP parameters optimization can be found in the SI.}
The corresponding learning curve is shown in Fig.~\ref{fig_learningcurve_molecule}(b). The improvement over a standard GPR scheme is systematic at any training set size, underlining the importance of automatically incorporating the $\mathcal{O}$(3)-covariance of the tensor at the scale of individual atomic environments.
Similarly to the case of GPR, the accuracy of predictions differs between tensor components. With 300 training points, the $\alpha_{xx}$ component presents the largest error (about 14\%), that reduces to only 6\% with 1500 training structures.
The best learning performance is again obtained for the $\alpha_{yz}$ component, for which the prediction error remarkably can be reduced to less than 2\%.  
As shown in Fig.~\ref{fig_Raman_GPR_molecule}, SA-GPR reproduces very accurately also the \textit{ab initio} Raman spectrum with only 300 training points.
Obviously, as reflected by the learning curves, increasing the amount of points reduces this error even further, as exemplified in Fig.~S4 of the SI, where the machine-learned and \textit{ab initio} spectra are virtually indistinguishable.

\subsection{Paracetamol crystal}

We now turn our attention to the first crystalline form of Paracetamol, containing four individual Paracetamol molecules per unit cell, as shown in Fig.~\ref{fig_paracetamol} (b).

\subsubsection{Direct approach}

Since we are now dealing with a periodic system, we first build a supercell corresponding to the appropriate crystal structure. Then, the AD representation is constructed following the same procedure discussed in the previous section. A value of $\sigma = 40$ has been selected to build the Gaussian kernel,  and the regularization parameter has been set to $\eta=10^{-4}$ by CV optimization.
The three-dimensional density field was evaluated within a box of $12 \times 14 \times 20$~\AA$^3$, using a grid spacing of $d\Br=0.75$~\AA\footnote{No noticeable improvement was observed when using a finer grid with $d\Br=0.5$~\AA.}.
For SA-GPR, the $\lambda$-SOAP  kernels were constructed using the same parameters as before.

\begin{figure}[htbp]
  \centering
  \includegraphics[width=1.0\columnwidth]{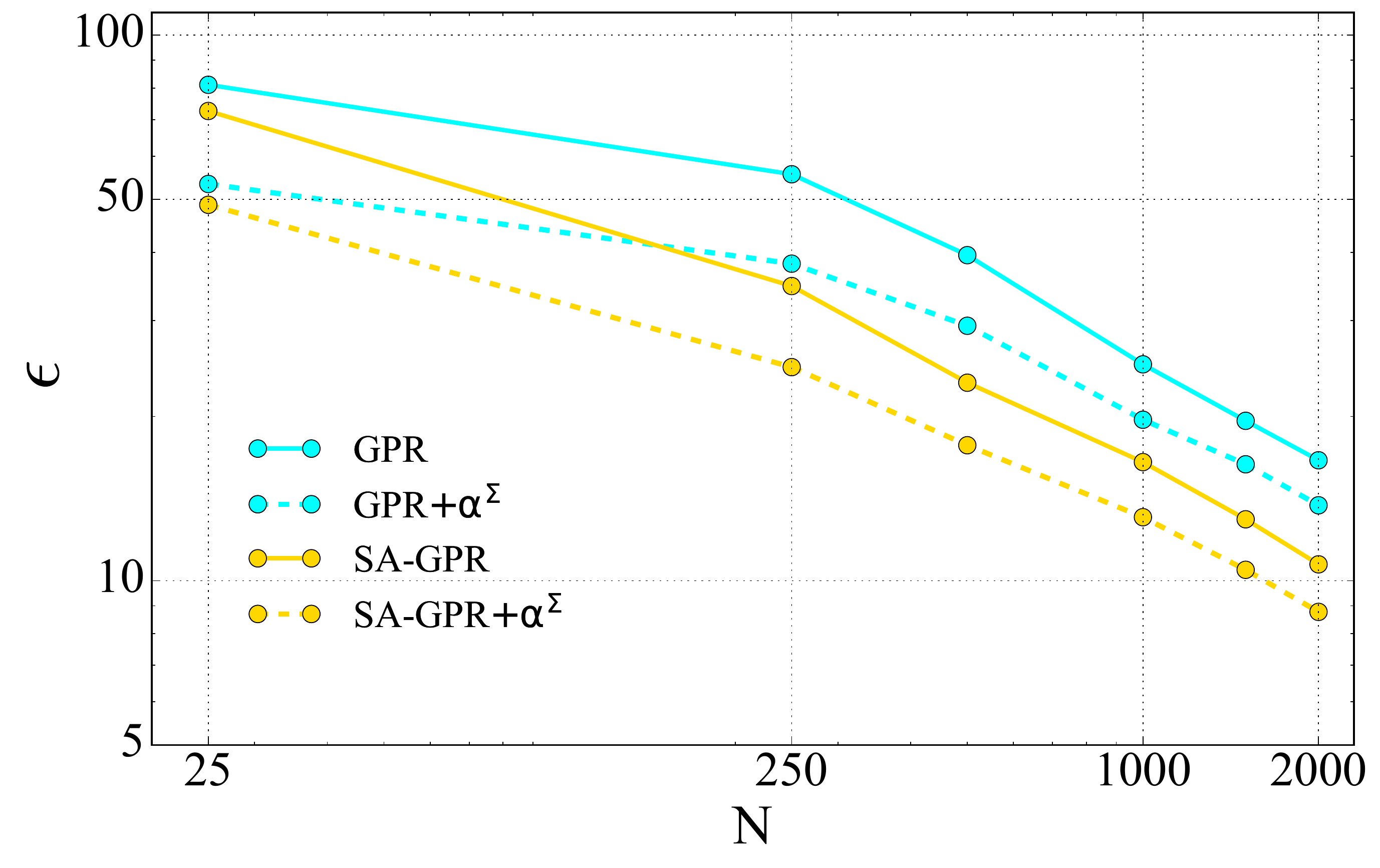}
  \caption{Learning curves for the crystal polarizability tensor on an NVT trajectory, using different approaches. Here the mean error over all components is represented. Including molecular polarizability greatly improves the model, both with GPR and SA-GPR, especially with few training data.}
 \label{fig_learningcurve_crystal}
\end{figure}

The training set is built by considering a random
selection of 2500 configurations extracted from a NVT trajectory. A full NVE trajectory is once again considered to test the quality of the predicted Raman spectrum. Learning curves for both regression models are shown in Fig.~\ref{fig_learningcurve_crystal} (solid lines).
When comparing the two methods, we always use the same configurations in both cases.

We observe that both the learning capability of GPR and SA-GPR do not reach saturation when increasing the number of training data, going from 81\% (respectively 73\% with SA-GPR) of error with 25 training points to 17\% (respectively 11\%) with 2000 of them;
Again, making use of a kernel that is built on a symmetry-adapted comparison between local environments brings a substantial improvement.
Overall, however, for the same amount of training points, the errors are much (typically between two and three times) larger than for the monomer case.

\subsubsection{Incorporating molecular polarizability}

To improve our results, we refined our models by using the predictions of the non-interacting monomers included in the molecular crystal\rev{, as we explain in the following}.

Suppose we have a molecular crystal made of $N_{\t{mol}}$ molecular units (in the case of Paracetamol I, $N_{\t{mol}}=4$, while $N_{\t{mol}}=8$ for Paracetamol II).
Since we have already learnt the polarizability tensors of the individual molecules, we can grasp most of the variability of the polarizability tensor of the crystal by summing up the predictions for individual monomers.
\rev{Equation}~\ref{eq_GPR} is modified according to
\begin{equation}
 \alpha^{\t{ML}}_{\gamma\delta}(\mathcal A)=\bar{\alpha}_{\gamma\delta}^{\t{ai,crys}}+\alpha^{\Sigma}_{\gamma\delta}(\mathcal A)-\bar{\alpha}^{\Sigma}_{\gamma\delta}+\sum_{j=1}^{N} w_{j} k(\mathcal A,\mathcal A_j) \,,
 \label{eq_GPR_molpol}
\end{equation}
where $\boldsymbol{\alpha}^\Sigma$ denotes the sum of the molecular polarizability tensors\rev{ and $\bar{\alpha}_{\gamma\delta}^{\t{ai,crys}}$ is the average of the \textit{ab initio} polarizability tensors of the full crystal over the training set}.
Specific details about the procedure are explained in \ref{sec:GPR_rot}.
An analogous expression to Eq. \ref{eq_GPR_molpol} is obtained for SA-GPR.

Figure ~\ref{fig_learningcurve_crystal} shows the advantage of using the molecular baseline for the regression models (dashed lines).
The improvement is most noticeable for models based on few training points. Molecular baselining leads to a large decrease in error of about 25\% for both GPR and SA-GPR. Upon increasing the number of training points, the difference diminishes, but an improvement remains visible.
It is worth noticing that, from 250 training structures on, the direct application of SA-GPR (without any molecular baseline) performs better than the GPR scheme with the baseline.
Note that the prediction accuracy can be improved even further if one scales the molecular polarizability tensors so that their average matches that of the full crystal, as illustrated in the SI, Fig.~S\rev{8}.

\begin{figure*}[htbp]
  \centering
  \includegraphics[width=0.9\textwidth]{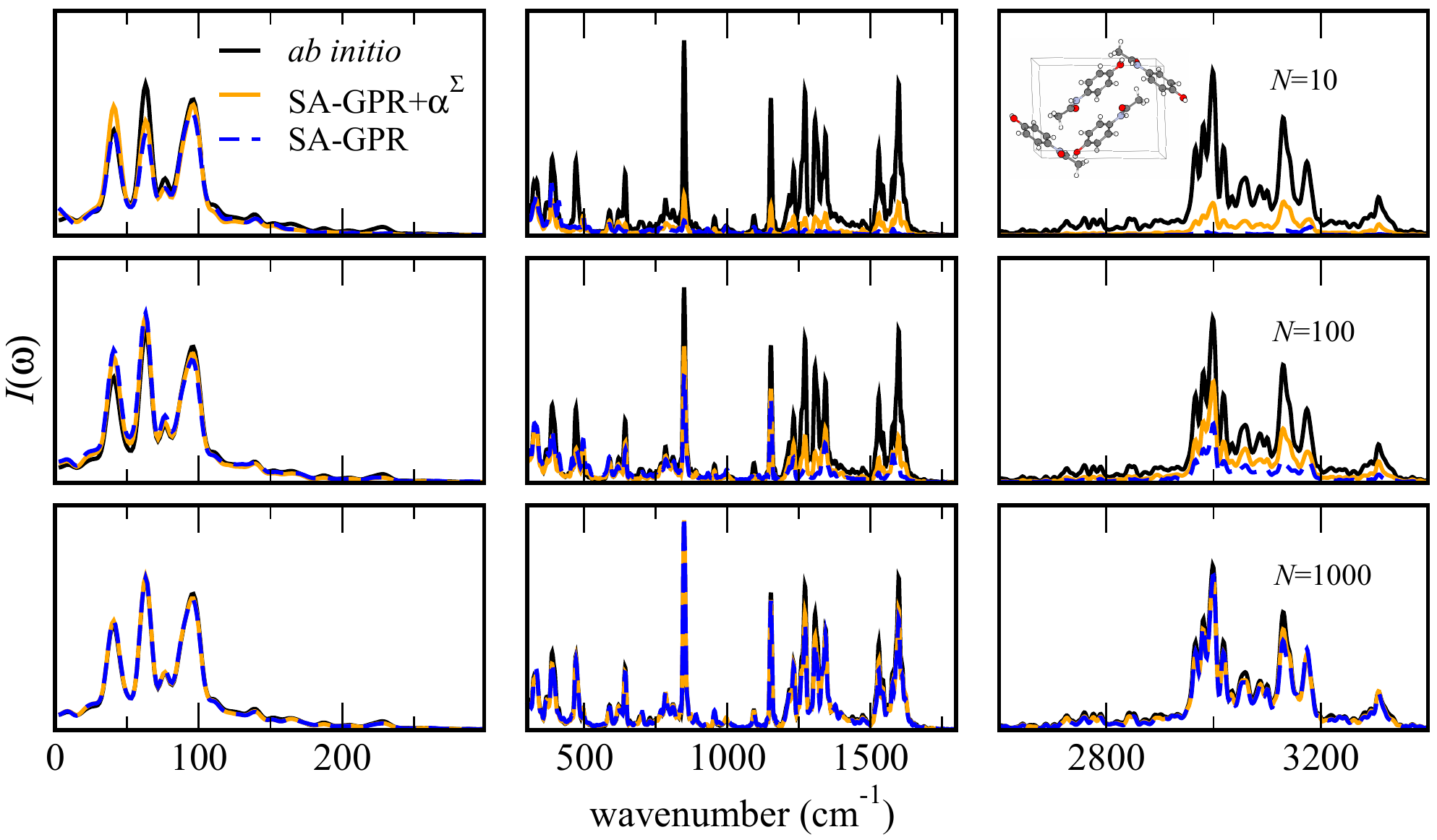}
  \caption{Raman spectrum of paracetamol I computed from a NVE trajectory, using either directly $\lambda$-SOAP SA-GPR, or augmenting this description by baselining it with molecular polarizabilities. The training was performed on NVT structures, the number of which is indicated in each row, while the spectrum was computed over 15000 consecutive NVE configurations.}
  \label{fig_crystal_raman_evol}
\end{figure*}

Figure~\ref{fig_crystal_raman_evol} shows the effect of the baselining procedure on the predicted SA-GPR Raman spectrum, when one increases the amount of training points.
Several observations can be made.
First, just like for the monomer, high frequencies require more training points to be reproduced, be it with a direct prediction or by including molecular polarizability tensors.
Second, including $\boldsymbol{\alpha}^{\Sigma}$ greatly enhances the spectrum intensity accuracy when few training points are used. This is especially true at high frequencies, where the improved model already gives the right structure of the peaks, albeit not with the right intensity, while the direct learning does not show any peak in this region.
Overall, the predicted spectrum is extremely well reproduced when employing enough training points.

\begin{figure*}[htbp]
  \centering
  \includegraphics[width=0.9\textwidth]{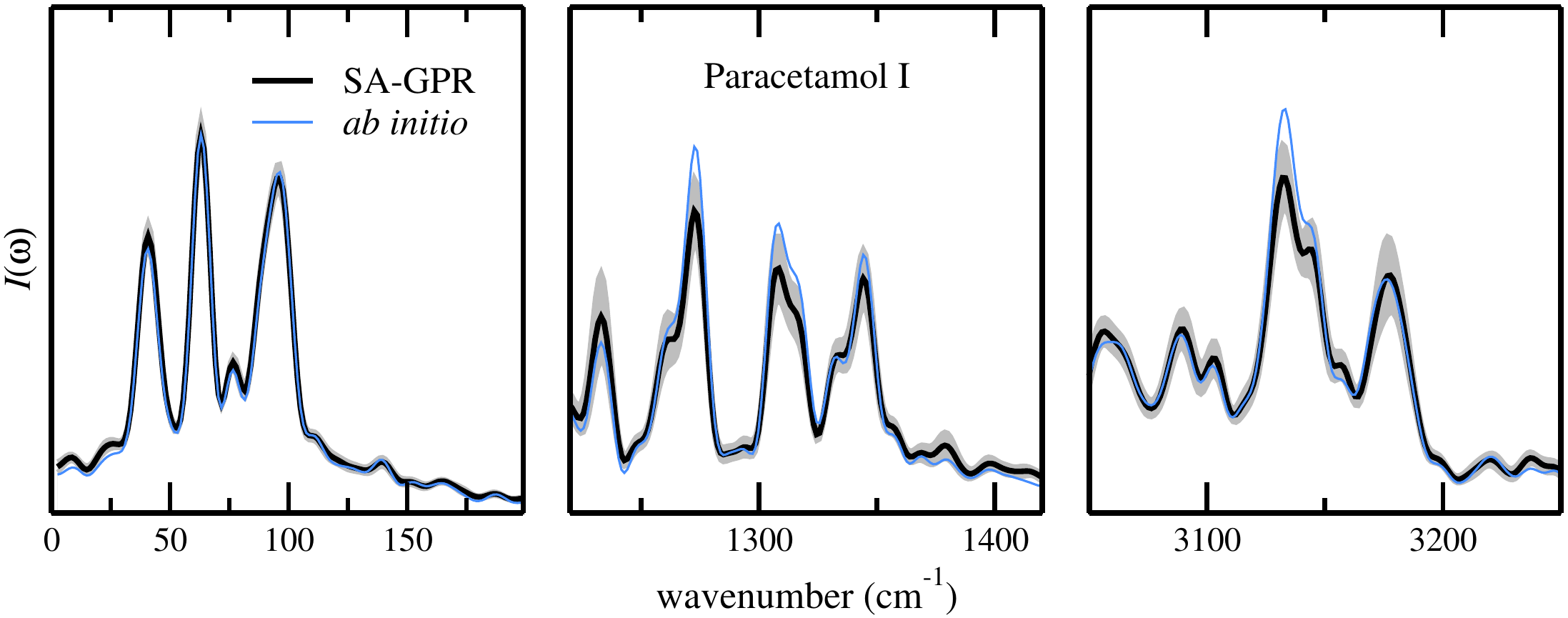}
  \caption{(\emph{black line}) Raman spectrum prediction of paracetamol form-I averaged over 16 different training models. Each training model is obtained by a random subselection of 2000 configurations over a total of 2500. (\emph{shaded area}) Standard deviation of the predicted spectra over the 16 models, \rev{calibrated with a likelihood maximization procedure described in Section \ref{subsec:errors}}. (\emph{blue line}) Reference \textit{ab initio} Raman spectrum.}
  \label{fig_crystal_sagpr_error}
\end{figure*}

Figure ~\ref{fig_crystal_sagpr_error} shows the predicted Raman spectrum and the corresponding estimated error. In this case, different learning models have been first defined by considering 16 random subselections, each of them made of 80\% of the training dataset. 
Then, for each of these learning models, the polarizabilities of the full NVE trajectory have been predicted and the associated Raman spectra have been computed. We then estimated the standard deviation of these predictions according to the procedure detailed in Sec.~\ref{subsec:errors}.
We find that in this case the estimated variance has to be increased by roughly an order of magnitude, i.e., $\nu^2 = 10.9$. 
One can observe that the excellent agreement between the reference and predicted spectrum at low frequencies is consistent with a negligible estimated error, while larger discrepancies and error bars can be observed in the high-frequency domain. 

\rev{
Results this far show that, even for a relatively rigid molecular system, incorporating symmetries and learning all the components of the polarizability tensor in a covariant fashion can improve the accuracy and the efficiency of these ML schemes. As anticipated, an additional advantage of SA-GPR is that it is based on local atomic environments. This results in a greater transferability, as we will discuss in the following section.
}

\section{Extrapolation on other polymorphic forms}
\label{sec:transfer}

Within the $\lambda$-SOAP formalism, the polarizability of the system is effectively decomposed in local atom-centered contributions that are summed in order to obtain the final predicted value of $\boldsymbol{\alpha}$\rev{, making it possible to model the susceptibility of a molecule or a crystal through the definition of effective atomic polarizabilities.}
This implies that the information is learned at the local scale and, as such, can be transferred across systems that share a similar chemical nature. 
In the case of paracetamol polymorphs, one can think of predicting the polarizability of the the form II crystal (Fig.~\ref{fig_paracetamol} (c)) with the model trained on form I only. Since different polymorphic forms are mainly distinguished by the different intermolecular interactions, major difficulties in this extrapolation procedure are expected to be associated with the low-frequency (intermolecular) modes of the molecular crystal. 
To put this idea to the test, we used the SA-GPR trained on 2000 structures of form I to make predictions for a NVE trajectory of form II. We used an ensemble of 16 subsampled models to estimate uncertainty.
As expected, we observed that the rather large error in the prediction of the polarizability tensor is mostly associated to a small offset in the time series of some of the polarizability components (detailed in the SI, Fig.~S10), that however does not have a substantial impact on the Raman spectrum. 

\begin{figure*}[htbp]
  \centering
  \includegraphics[width=0.9\textwidth]{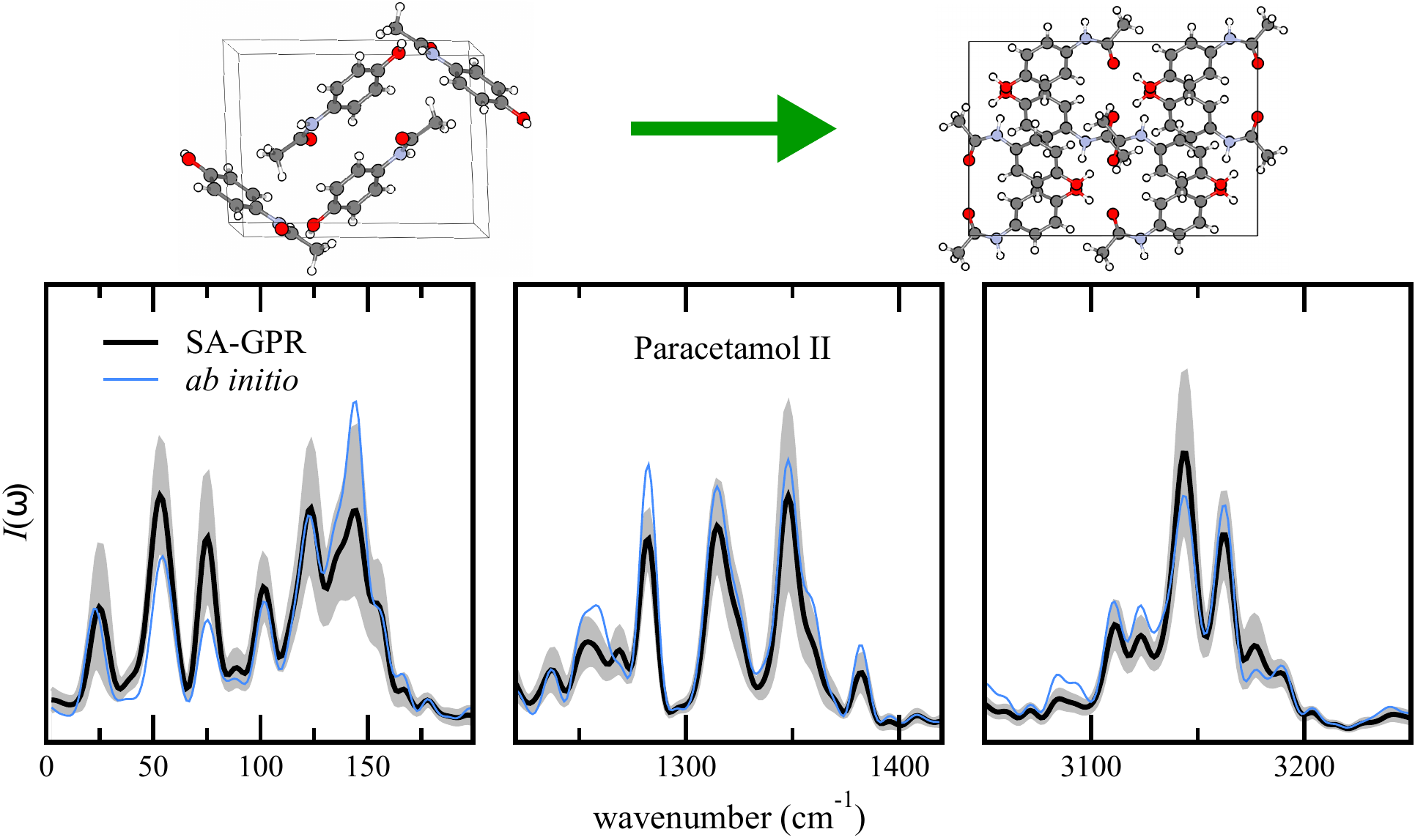}
  \caption{(\emph{black line}) average Raman spectrum prediction of paracetamol form-II associated with the same 16 training models already used for the prediction of paracetamol form-I. (\emph{shaded area}) standard deviation of the predicted spectra over the 16 models\rev{, calibrated with a maximum likelihood procedure described in Section \ref{subsec:errors}}. (\emph{blue line}) reference \textit{ab initio} Raman spectrum.}
  \label{fig_crystal_sagprII}
\end{figure*}

As shown in Fig.~\ref{fig_crystal_sagprII}, the general lineshape is excellent, and all the main features of the \textit{ab initio} spectrum are reproduced, even though few  discrepancies in terms of intensity can be observed, and the error in the intensities is overall higher than for the direct prediction of the first polymorph. 
We underline the difference in behaviour in contrast to Fig.~\ref{fig_crystal_sagpr_error}: now, high frequencies are better described and errors are more pronounced at low frequencies.
\rev{This suggests that the model can reproduce accurately changes in the polarizability associated with intra-molecular vibrations, but is less accurate in predicting low-frequency components that are specific to the molecular packing of form II, which is not represented in the training set for form I. 
We also observe that the discrepancy between predictions and \emph{ab initio} spectrum is reflected accurately in the estimated uncertainty, that can therefore be used as a reliable measure of the accuracy of the model also in an extrapolative regime. }

\section{Conclusions}

In this work, we proposed GPR models to predict vibrational Raman spectra, based on learning
polarizability and susceptibility tensors obtained from density-functional perturbation theory. As an example, we applied our methodology to predict anharmonic Raman spectra of the Paracetamol molecule and two polymorphs of the Paracetamol crystal. The methodology also works for simpler harmonic Raman intensities.
The use of an ensemble of models to estimate the uncertainty in the polarizability tensors allows us to propagate the error estimation from the ML prediction of the polarizability tensors to the vibrational spectra, by generating an ensemble of spectra out of which it is simple to compute frequency-dependent confidence intervals.

We showed that for the molecule a standard GPR scheme that takes as input a nuclear density representation on a 3D grid works extremely well and enables one to reproduce Raman spectra almost perfectly with a low number of training points.
For the crystal, such a scheme, albeit possible, is more difficult to apply for several reasons: The difficulty to compare crystal structures with different unit cell sizes, the redundancy of information contained in a fixed grid-based representation, and the increase of the number of  grid points with system size. 
We have shown that a more effective solution is to use a symmetry-adapted GPR scheme, used here in combination with the $\lambda$-SOAP representation \cite{Grisafi_SAGPR}. Such a scheme yields accurate predictions for molecules and crystals, due to its capability to better capture the local structural information in a covariant fashion. Moreover, since $\lambda$-SOAP is a local representation, it is easy to treat larger systems sizes and even transfer models to other polymorphs. We have shown this transferability by successfully predicting the Raman spectrum of Paracetamol II with a model trained only on Paracetamol form I.
This suggests the possibility of predicting Raman spectra of any polymorphic form, as long as a model trained for one of them is available.
In addition, for all models presented, we observe a considerable improvement when using previously-trained GPR models for the molecular units as a baseline for the crystal prediction, thus reducing the amount of more costly condensed-phase calculations that must be performed to train the bulk model. In a similar manner, it is also straightforward to extend this framework to other ensembles (e.g. NPT) or path-integral molecular dynamics simulations, which include the quantum nature of the nuclei.

The models we presented regard the electronic electric-field response properties, and can be extended to dipoles and higher-order responses. They can thus be seamlessly combined with empirical potentials or other machine-learned potentials that give access to forces.
This presents an alternative route to including the training of such quantities directly into these potentials~\cite{Gastegger_ChemSci_2017}, which can present a higher level of complexity.  
Finally, we remark that even though we applied our framework to polarizabilities and Raman spectroscopy, applying it to any other kind of spectroscopy, like infrared, sum-frequency generation, etc., would be straightforward, as long as reference electronic-structure data is available.

\section{Acknowledgments}
The authors acknowledge funding from the Max Planck-EPFL center. MC acknowledges support from the European Research Council (Horizon 2020 Grant Agreement No. 677013-HBMAP).

\appendix

\section{Molecular baselining with GPR}
\label{sec:GPR_rot}

\begin{figure}[htbp]
  \centering
  \includegraphics[width=1.0\columnwidth]{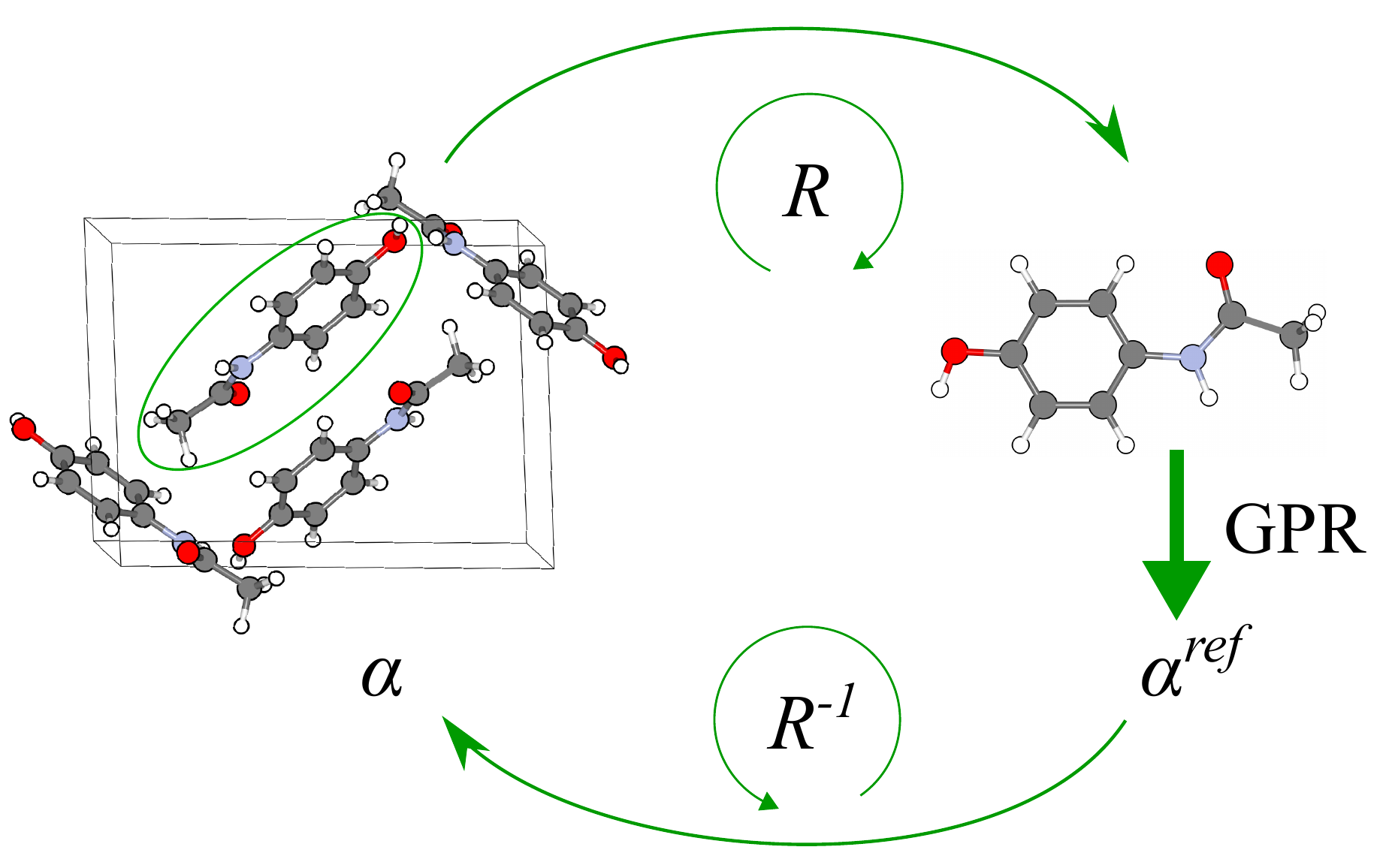}
  \caption{Schematic explanation of the molecular baselining process for the \rev{standard} GPR approach. An individual molecule of the crystal is first rotated by a rotation matrix $R$ to match the alignment of a reference structure, on top of which we calculate the polarizability \rev{with GPR} using Eq.~\ref{eq_GPR}, and we then rotate back. We repeat this process for every molecule in the unit cell, and sum the resulting polarizability tensors to obtain $\alpha^\Sigma$.}
  \label{fig_GPR_mol_baseline}
\end{figure}

When building a model to predict the polarizability tensor of the crystal from its individual molecular components within the GPR and AD representation scheme, one should consider that each molecule is oriented in a specific direction that differs from the one we trained the molecule on. In order for the regression model to recognize the orientation at hand, we first need to find the relationship between the orientation of the molecules in the crystal and the \rev{reference} one used in the molecular GPR procedure described in \rev{section} \ref{subsec:GPR}. The scheme is depicted in Fig.~\ref{fig_GPR_mol_baseline}.
%
%

Each geometry $G_{im}$ corresponding to the $i$th molecule \rev{in the crystal} of the $m$th structure \rev{of the set} is thus rotated by a rotation matrix $R_{im}$ as
\begin{equation}
    G^{\t{ref}}_{im}=R_{im} G_{im} \,,
\end{equation}
where $i=1\cdots N_{\t{mol}}$\rev{, with $N_{\t{mol}}$ the number of molecular blocks per unit cell}.
Finally, once a molecular polarizability $\boldsymbol{\alpha}^{\t{ref\rev{,mol}}}_i$ is predicted, we rotate the tensor back to its original orientation inside the crystal, i.e., 
\begin{equation}
    \boldsymbol{\alpha}_{i,m}^{\t{mol}}=R^{-1}_{im}\boldsymbol{\alpha}_{i,m}^{\t{ref,mol}} R_{im} \,.
\end{equation}
Having defined the sum of molecular polarizabilities $\boldsymbol{\alpha}^{\Sigma}$ as
\begin{equation}
    \boldsymbol{\alpha}^{\Sigma}=\sum_{i=1}^{N_{\t{mol}}} \boldsymbol{\alpha}_{i}^{\t{mol}}\,,
\end{equation}
we consider the regression target 
\begin{equation}
   \Delta\alpha_{\gamma\delta}\rev{(\mathcal A)}= \alpha^{\t{crys}}_{\gamma\delta}\rev{(\mathcal A)}-\bar{\alpha}^{\t{crys}}_{\gamma\delta} - (\alpha^{\Sigma}_{\gamma\delta}\rev{(\mathcal A)}-\bar{\alpha}^{\Sigma}_{\gamma\delta})\ , 
\end{equation}
where, once again, the bar denotes the average over the training set. 
\rev{Finally, we modify accordingly} equation~\ref{eq_GPR}, \rev{which} becomes \rev{(see also Eq.~\ref{eq_GPR_molpol})}
\begin{equation}
 \alpha^{\t{ML}}_{\gamma\delta}(\mathcal A)=\bar{\alpha}_{\gamma\delta}^{\t{ai,crys}}+\alpha^{\Sigma}_{\gamma\delta}(\mathcal A)-\bar{\alpha}^{\Sigma}_{\gamma\delta}+\sum_{j=1}^{N} w_{j} k(\mathcal A,\mathcal A_j) \,.
 \label{eq_GPR_molpol_appendix}
\end{equation}
\rev{Note that the weights in the previous equation are different than the ones in Eq.~\ref{eq_GPR}, as this time the regression target $\Delta\alpha$ contains the molecular polarizability tensors.}

An analogous expression is obtained for SA-GPR, but the rotation and alignments previously described do not need to be carried out explicitly since the rotational covariance of the tensor is built in the structure of the method.

\bibliographystyle{iopart-num}
\bibliography{main, refs} 

\end{document}


\title[SI: Gaussian Process Regression for Raman Spectra]{Supplemental Information: Using Gaussian Process Regression to Simulate the Vibrational Raman Spectra of Molecular Crystals}
\author{Nathaniel Raimbault$^{1}$\footnote{These authors contributed equally to the manuscript.}, Andrea Grisafi$^{2}\ddagger$, Michele Ceriotti$^2$, Mariana Rossi$^1$}
\address{$^1$ Fritz Haber Institute of the Max Planck Society, Faradayweg 4-6, 14195 Berlin, Germany}
\address{$^2$ \'Ecole Polytechnique F\'ed\'erale de Lausanne, Route Cantonale, 1015 Lausanne, Switzerland}
\ead{michele.ceriotti@epfl.ch and rossi@fhi-berlin.mpg.de}

%
%
\submitto{\NJP}
%
%
\ioptwocol

\section{Optimizing hyperparameters}
\label{sec:hyper_gpr}

\begin{figure}[hbtp]
  \centering
  \includegraphics[width=0.5\columnwidth]{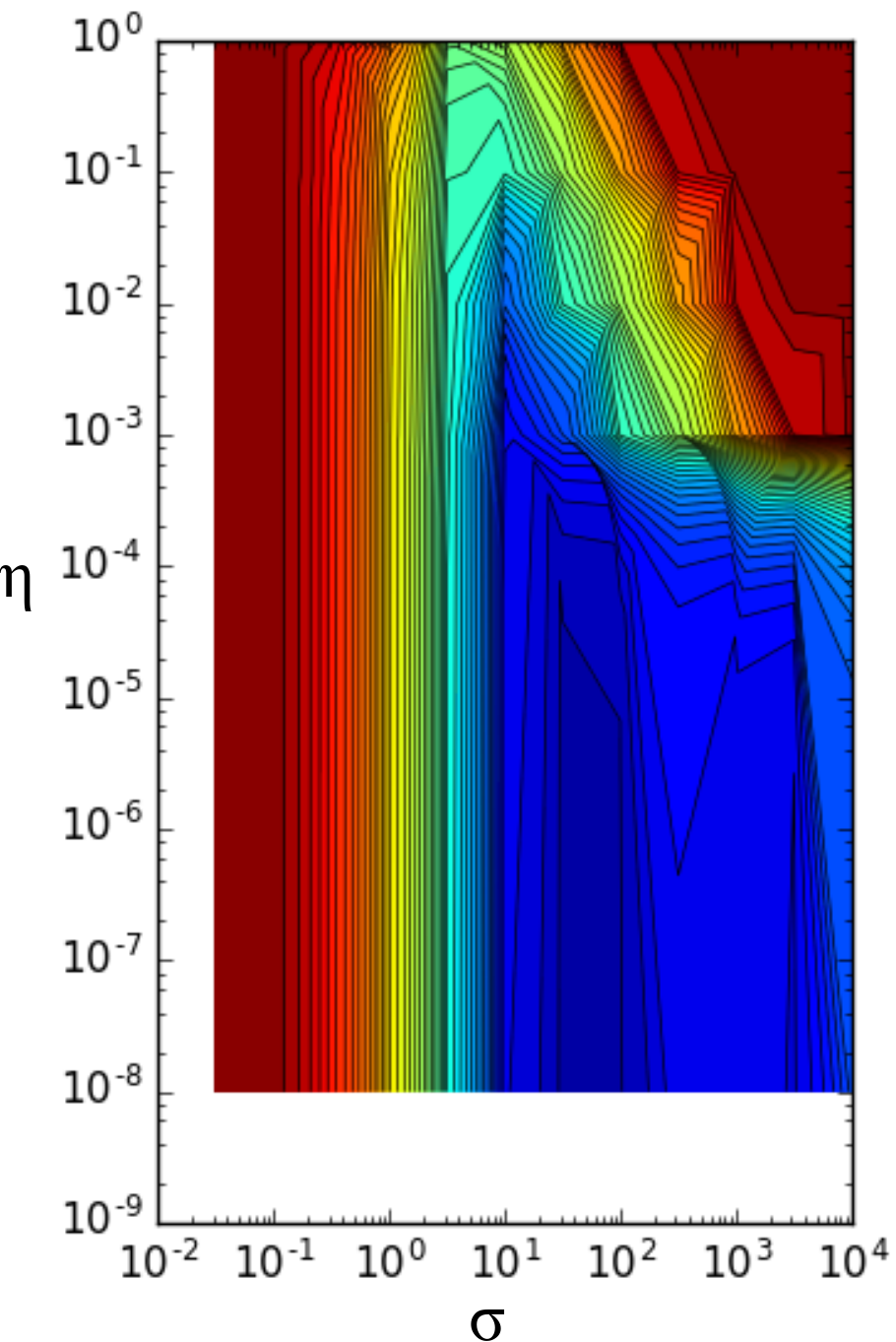}
  \caption{Contour plot of $\epsilon_{xx}$, using GPR with a density grid representation, with 300 training points and 100 test points. Blue (red) indicates smaller (larger) errors. The contour actually shifts to the left when increasing the number of training points.}
  \label{fig_contour_hyperparameters}
\end{figure}

In order to find the optimal hyperparameters $\sigma$ and $\eta$ used in GPR, we have used a cross-validation method. 
We picked a set of structures (here by farthest point sampling) from an NVT trajectory, which we then divided into a training and validation set. We then varied the hyperparameters and calculated the error (see Eq.~6) on the validation set in each case. 
In Fig.~\ref{fig_contour_hyperparameters} we show an example of a contour plot of $\epsilon_{xx}$ for Paracetamol I.
Very similar profiles are obtained for all polarizability components.
\rev{The values for $\sigma$ and $\eta$ given in the main text have been optimized using random subselections of 900 structures for the training and 300 for the validation set. 

A similar procedure has been carried out to optimize the radial cutoff $r_\t{cut}$ and the Gaussian width $\sigma$ entering the SOAP-kernel definitions. In this case, random subselections of 1500 configurations out of a total of 2000 has been chosen to train the model while the remaining 500 are selected as a validation set. Both for the monomer and the crystal case, we find that $\sigma=0.3$~{\AA} and $r_\t{cut}$=4.0~{\AA} yield the best learning performances. We multiplied the linear $\lambda$-SOAP kernel by a $\lambda=0$ scalar kernel with the same parameters, so as to obtain a non-linear kernel that captures higher-order structural correlations without sacrificing the covariant behavior of the model~\cite{Wilkins2019}. }

\section{Learning curves}
 \subsection{Atomic coordinates}
\begin{figure}[htbp]
   \includegraphics[width=1.0\columnwidth]{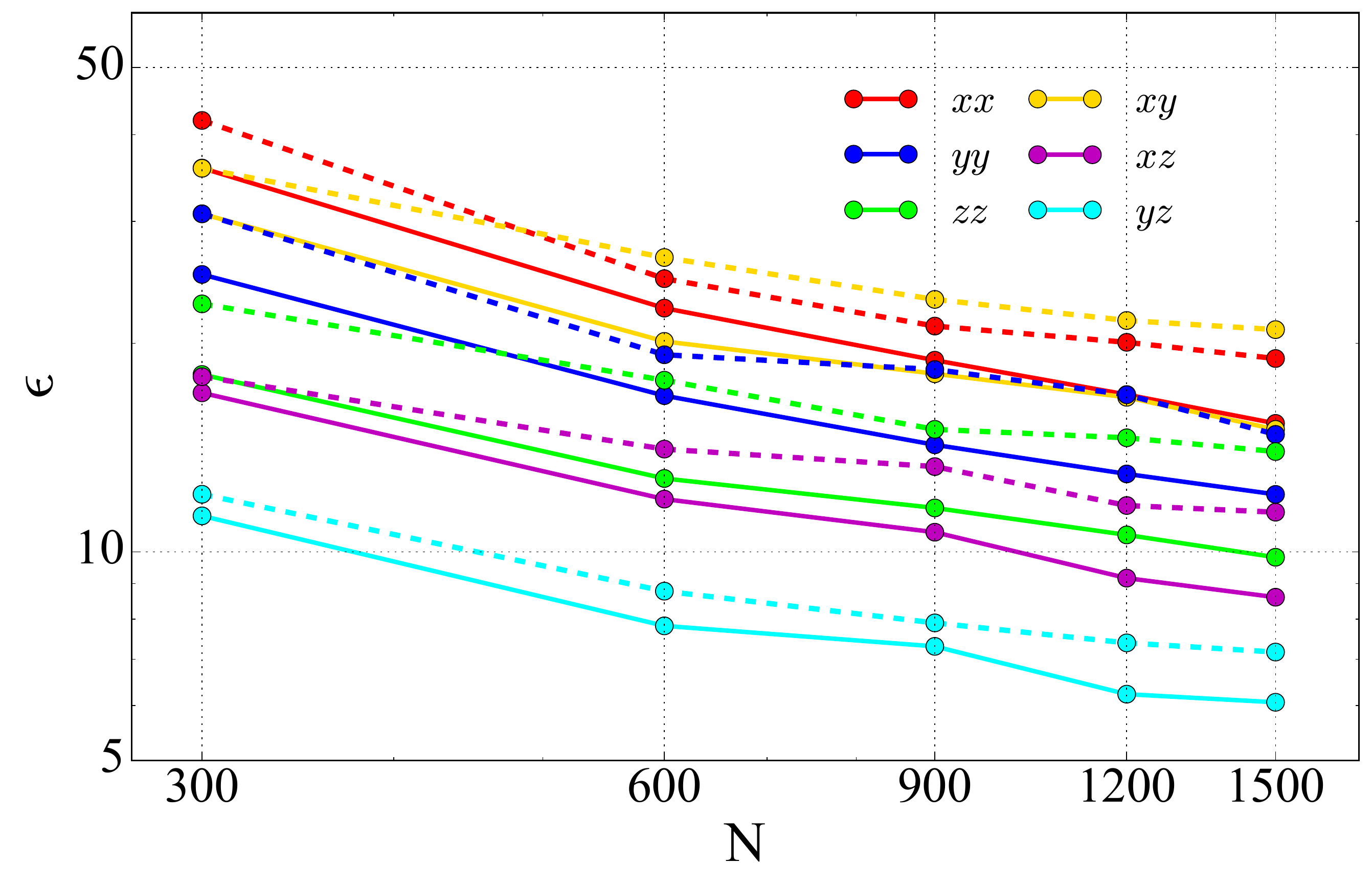}
  \caption{Learning curves for each component of the polarizability tensor of the paracetamol molecule, using GPR with raw atomic coordinates (dashed lines) and atomic densities (solid lines) as representations. The errors are given by Eq. 6 in the main text.}
 \label{fig_learningcurve_molecule_raw}
\end{figure}

In the main text, Gaussian process regression was systematically used with a density-grid representation. In this section we show the performance of a GPR approach based on unprocessed coordinates in Fig.~\ref{fig_learningcurve_molecule_raw}.
Note that representing the system through an atomic coordinates only entails using the cartesian components as the feature vector $\boldsymbol{u}(\mathcal{A})$, thus requiring an alignment to a reference structure.
We observe that the learning curve when simply taking raw coordinates to train the model shares a similar behaviour with the atomic density representation, but is systematically a little lower, meaning that it needs more training points to reach the same level of accuracy.
Therefore the simplest representation with unprocessed atomic coordinates was not considered in the rest of this article, where GPR is systematically employed in combination with atomic densities. 


%
\begin{figure*}[htbp]
  \centering
  \includegraphics[width=0.85\textwidth]{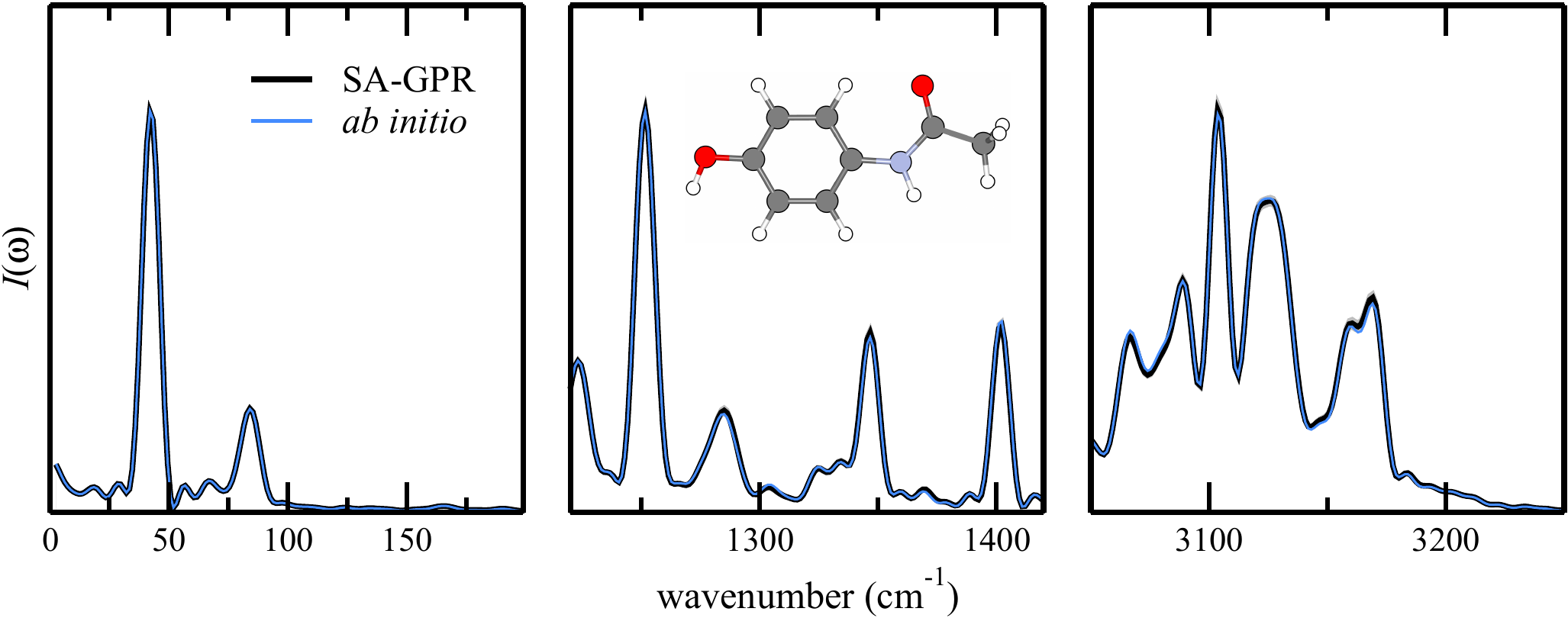}
  \caption{(\emph{black line}) Raman spectrum SA-GPR prediction of the Paracetamol molecule averaged over 16 different training models. Each training model is obtained by a random subselection of 2000 configurations over a total of 2500. (\emph{shaded area}) standard deviation of the predicted spectra over the 16 models. (\emph{blue line}) reference \textit{ab initio} Raman spectrum. }
  \label{fig_Raman_SAGPR_molecule}
\end{figure*}

\section{Fully converged Raman spectrum (molecule)}

To make the parallel with Figs~10 and 11 in the main text, we show in Fig.~\ref{fig_Raman_SAGPR_molecule} the Raman spectrum of the Paracetamol molecule, using SA-GPR trained on different training models, containing each 2000 structures.

We see that with this amount training points, SA-GPR reproduces exactly the \textit{ab initio} spectrum of the Paracetamol molecule;
At this level of accuracy, the error bars are barely visible, and low frequencies and high frequencies are equally well described.

Furthermore, comparison with Fig.~10 in particular highlights the difficulty in predicting a molecular crystal rather than one of its molecular constituents: many more training points are needed in the case of Paracetamol I to reach the same level of accuracy as that of the molecule.

\section{Variations in polarizability predictions}

\begin{figure}[htbp]
  \centering
  \includegraphics[width=1.0\columnwidth]{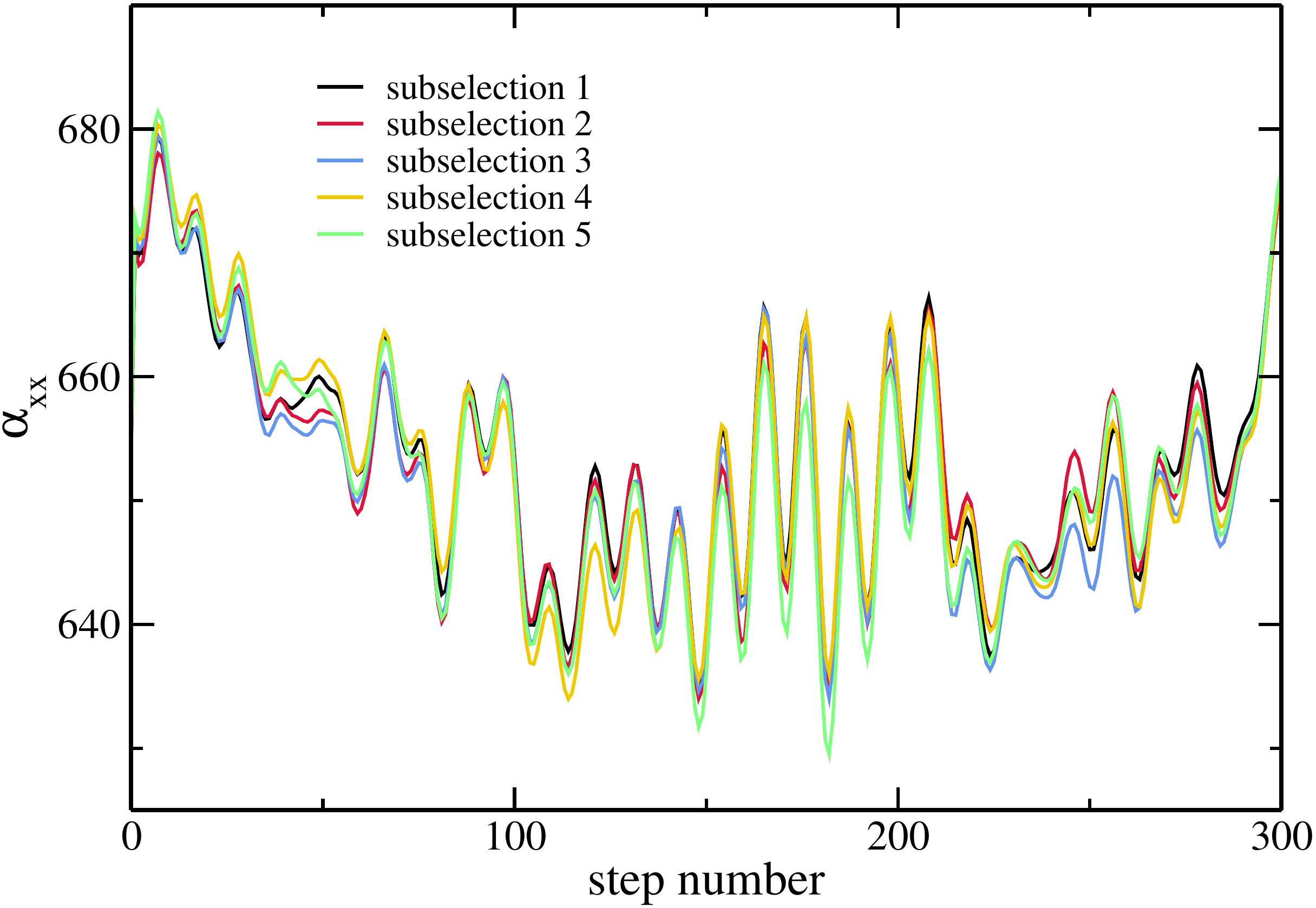}
  \caption{Predicted polarizability ($xx$ component) time series with GPR. The validation test is fixed in each case, while the we use 5 different subselections for the training set, the size of which is fixed to 2000 structures.}
  \label{fig_crystal_polar_variations}
\end{figure}

\begin{figure}[htbp]
  \centering
  \includegraphics[width=1.0\columnwidth]{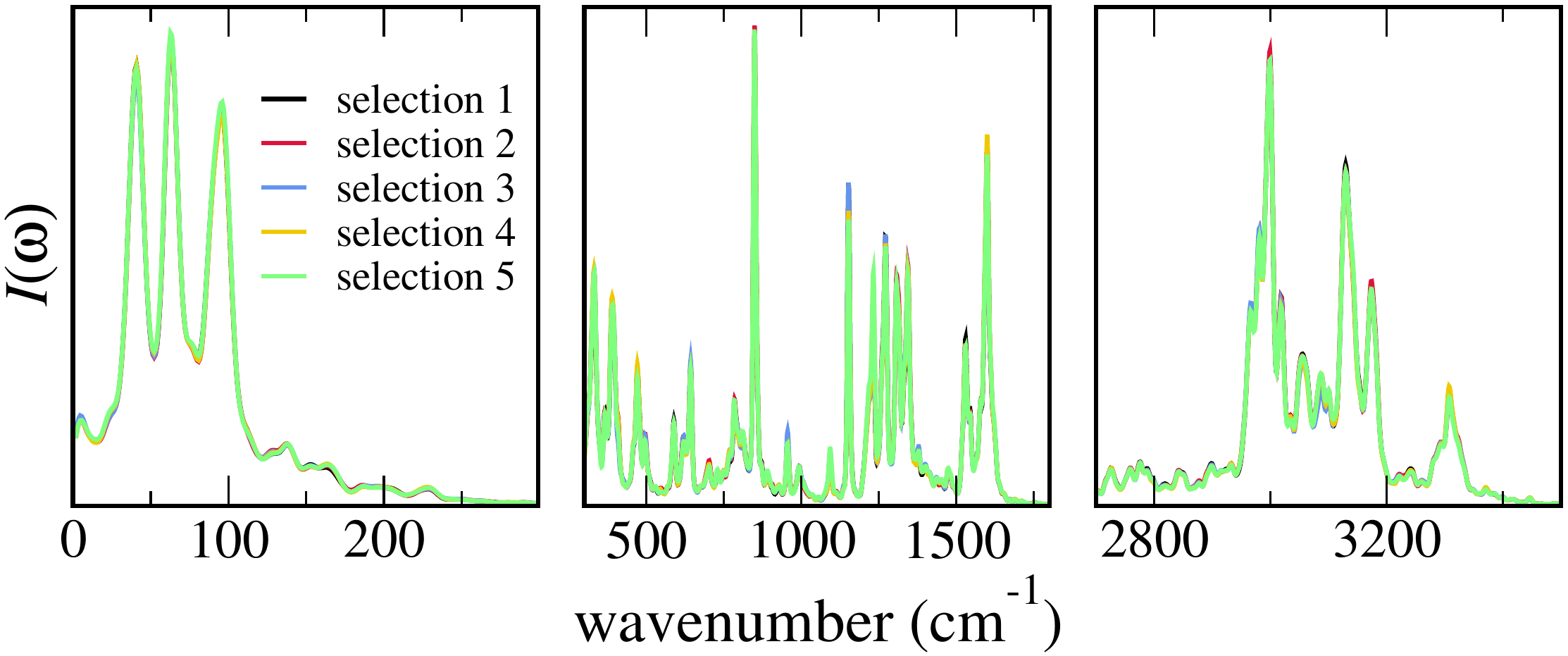}
  \caption{Predicted Raman spectrum of paracetamol I with GPR, using 5 different subselections of constant size (2000 structures) of training data.}
  \label{fig_crystal_Raman_variations}
\end{figure}

In the main text, we showed results that were obtained each time with a unique subselection of the training set. To show the dependence of our predicted polarizabilities on the choice of the training data, we chose a fixed validation set, but varied the selection of training data, keeping its size constant. Results are shown in Fig.~\ref{fig_crystal_polar_variations}.
While some substantial modifications are observed on a point-by-point comparison basis, the general shape and variations are extremely similar for each of the subselections. Consequently, the corresponding Raman spectrum is impacted very little by such variations, as we demonstrate in Fig.~\ref{fig_crystal_Raman_variations}.
It is thus usually sufficient to have a single model to predict new trajectories, as long as one uses enough training points.

\section{Molecules in crystal}

\begin{figure}[h]
  \centering
  \includegraphics[width=1.0\columnwidth]{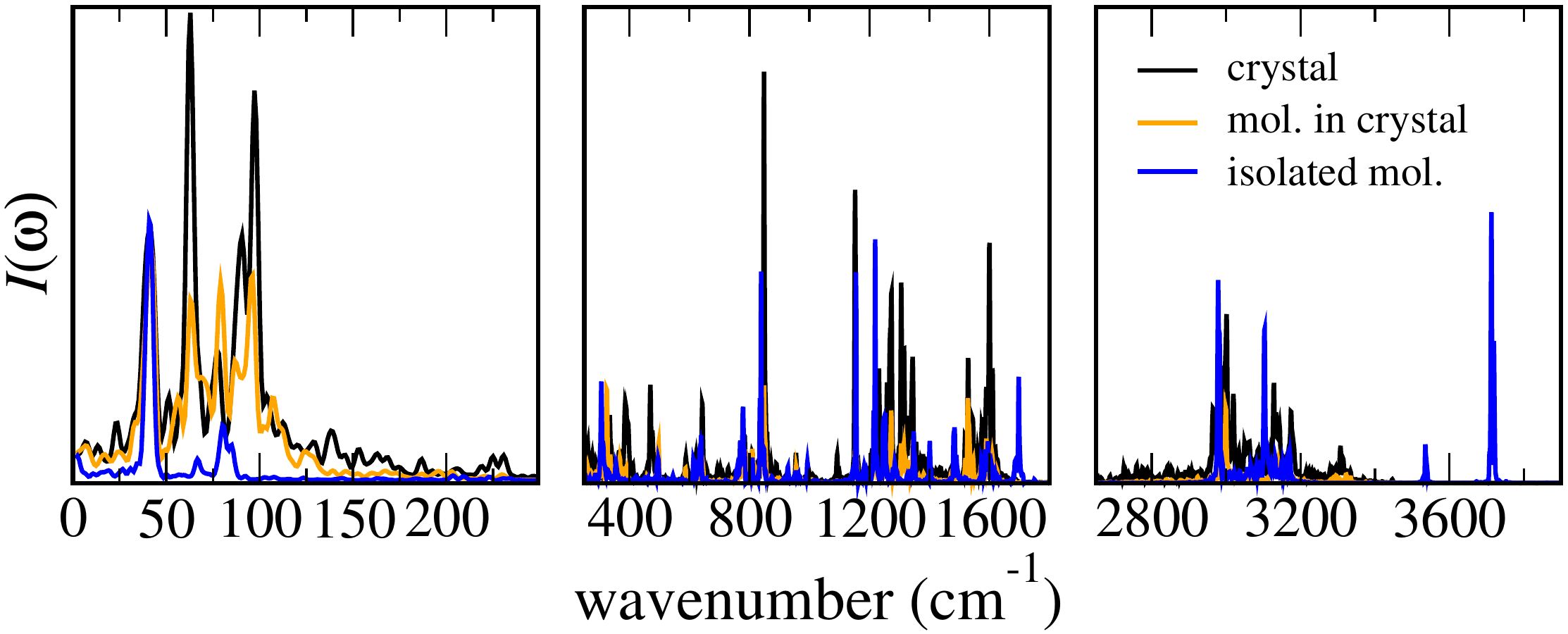}
  \caption{Raman spectrum of one interacting molecule of the crystal as compared to the isolated molecule and to the actual crystal spectrum.}
  \label{fig_crystal_mol}
\end{figure}

Let us have a look at the Raman spectrum of a single molecule as behaving inside the crystal, which we shall call "interacting molecule", and compare it to that of the isolated molecule, shown in Fig.~\ref{fig_crystal_mol}.
We see that the two spectra are very different. The very high-frequency peaks of the isolated molecules (O-H and N-H stretching motions) disappear in the case of the interacting molecule, and also the low-frequency range shows some important changes. Actually, the interacting-molecule spectrum looks much closer to the full crystal spectrum.

\begin{figure}[htbp]
  \centering
  \includegraphics[width=1.0\columnwidth]{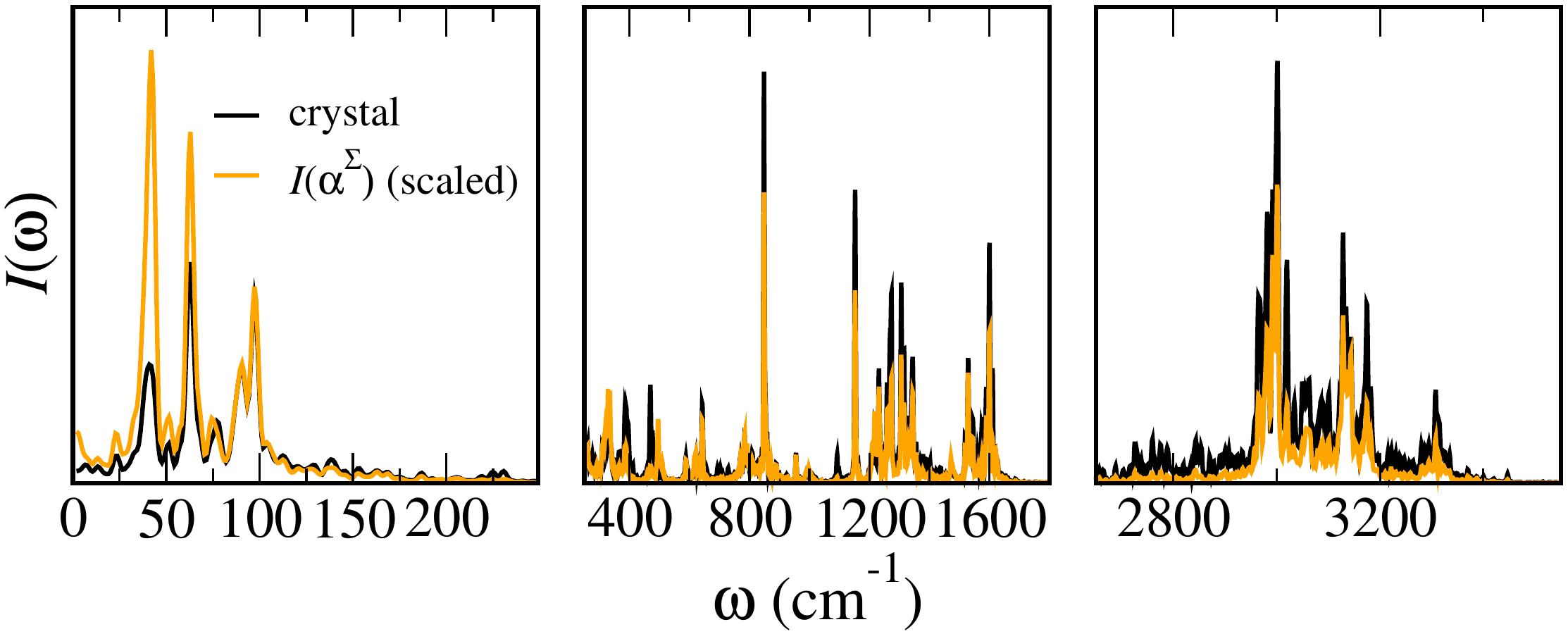}
  \caption{Raman spectrum computed directly from the sum of molecular polarizabilities (obtained with GPR) as compared to the actual crystal spectrum. No machine learning is used here for representing the crystal; However, note that the spectrum obtained from $\alpha^{\Sigma}$ has been arbitrarily scaled (different factor for each panel) so that both intensities are comparable for every frequency range.}
  \label{fig_crystal_frommolpol}
\end{figure}

It actually seems somewhat counter-intuitive that the molecular spectrum would differ that much at high frequency. Indeed, the higher-frequency vibrational modes should be governed by intramolecular vibrations, and in this regard are expected to be similar within the crystal. 
Interestingly now, if we compute directly the autocorrelation function of $\alpha^{\Sigma}$, i.e., the sum of molecular contributions, and compare it to the full crystal spectrum, we find a very good agreement, as shown on Fig.~\ref{fig_crystal_frommolpol}, provided that we apply \textit{a posteriori} a fixed scaling factor for each different frequency range.
We see in particular that the three main peaks of the low-frequency range are present in the molecular spectrum, although not with the right intensity.
We note that for the latter figure, we used Gaussian process regression for predicting the individual molecular polarizabilities, but we did not use any machine-learning scheme to predict directly the crystal polarizabilities. All the information coming from the interaction between the different molecules is thus absent, and still needs to be grasped by a machine-learning approach.

\begin{figure}[h]
  \centering
  \includegraphics[width=1.0\columnwidth]{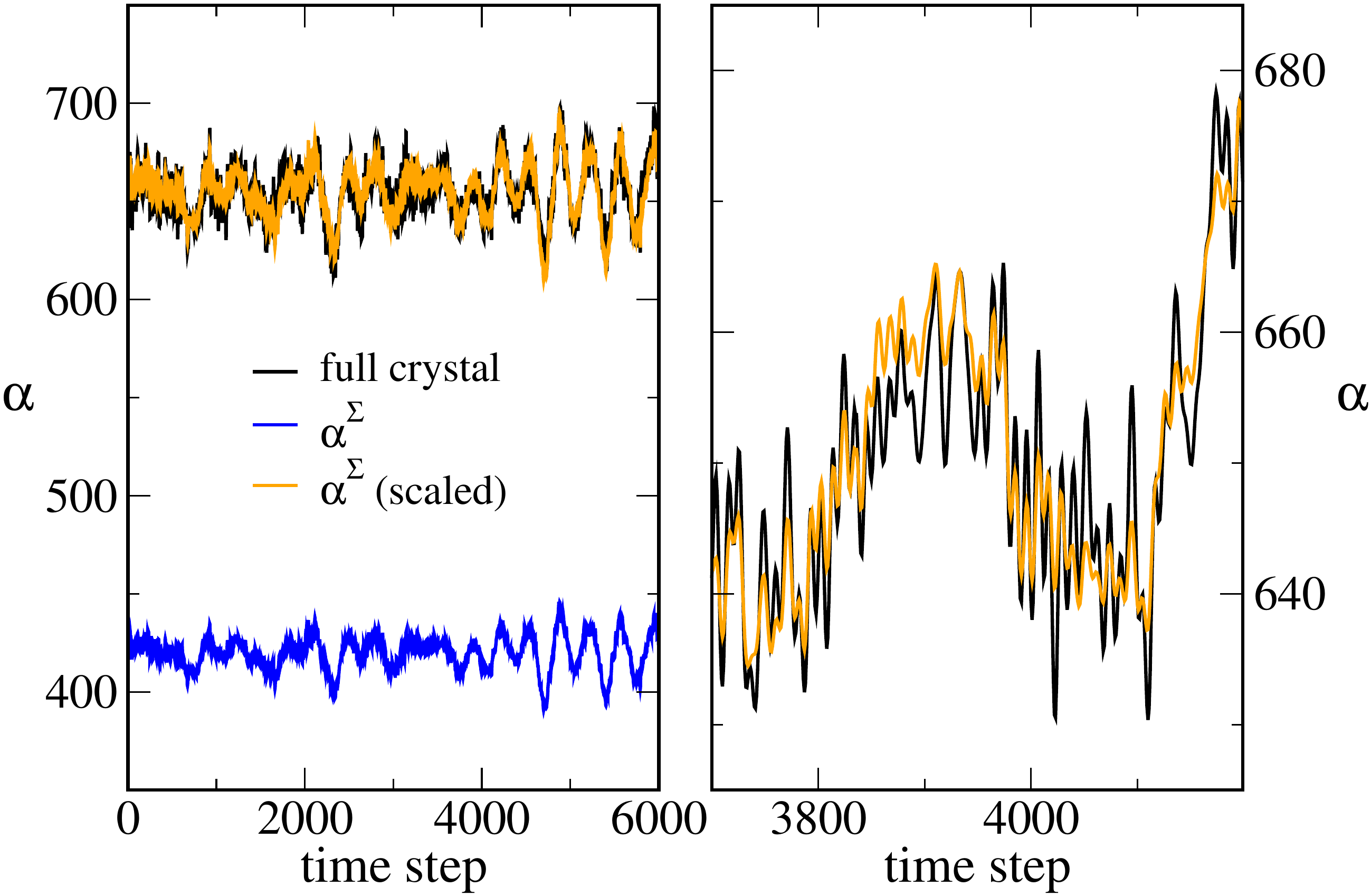}
  \caption{Time series of the full crystal polarizability, the molecular polarizability, and the scaled molecular polarizability. Only the $xx$ component is shown. An analogous behavior is observed for the other components.}
  \label{fig_crystal_molpol_series}
\end{figure}

As illustrated in Fig.~\ref{fig_crystal_molpol_series} for GPR, the prediction accuracy can be improved even further if one scales the molecular polarizability tensors so that their average matches that of the full crystal.
The variations of $\boldsymbol{\alpha}^{\Sigma}$ are very close to that of the full polarizability of the crystal, but both its absolute values and derivatives are considerably smaller. The scaled version reduces these differences. 

\section{Harmonic vibrational Raman spectra}

A more widely used possibility to simulate vibrational Raman spectra in an \textit{ab initio} manner is through the harmonic approximation, in which the potential energy is Taylor-expanded up to second order; Within this approximation, harmonic Raman intensities $I_H(\omega)$ are proportional to the derivatives of the polarizability tensor with respect to atomic displacements (see e.g., Refs.~\cite{Neugebauer_JCC2002,Veithen:2005gf,Prosandeev2005}):
\begin{eqnarray}
I_{H}(\omega) & \propto \frac{1}{\omega(1-e^{-\frac{\hbar\omega}{k_BT}})} \frac{1}{30} (10 G_p^{(0)}+7 G_p^{(2)}) \,, \label{eq:harm-raman} \\
\nonumber G_p^{(0)} & =\frac{1}{3}\left[ ({\alpha}_{xx}')_p + ({\alpha}_{yy}')_p + ({\alpha}_{zz}')_p \right]^2 \\
\nonumber G_p^{(2)} & =\frac{1}{2} \left[(2 {\alpha}_{xy}')_p^2 + (2 {\alpha}_{xz}')_p^2 + (2 {\alpha}_{yz}')_p^2\right] + \\ \nonumber & + \frac{1}{3} \{ \left[({\alpha}_{xx}')_p-({\alpha}_{yy}')_p\right]^2 +  \\ \nonumber & + \left[({\alpha}_{xx}')_p-({\alpha}_{zz}')_p\right]^2 + \left[({\alpha}_{yy}')_p-({\alpha}_{zz}')_p\right]^2 \}
\end{eqnarray}
where $({\alpha}_{ij}')_p=\left(\partial {\alpha}_{ij}/\partial Q_p\right)_0$ is the derivative of the $ij$ component of the polarizability with respect to the displacement of normal mode $Q_p$.
In practice, we evaluate these derivatives by finite differences: we make $6n$ (forward and backward, $n$ being the number of atoms per unit cell) small nuclear displacements in the unit cell around the equilibrium position, on top of which we compute the polarizability tensor with DFPT. 

Because in the harmonic approximation, every calculation takes place close to equilibrium, the standard deviation of the polarizability in that case will be very small. The challenge to predict the latter with machine learning is thus slightly different than in the previous case, where polarizabilities can experience large variations. Here the model will have to be sensitive to very small symmetric alterations in the structure.

\begin{figure}[htbp]
  \centering
  \includegraphics[width=1.0\columnwidth]{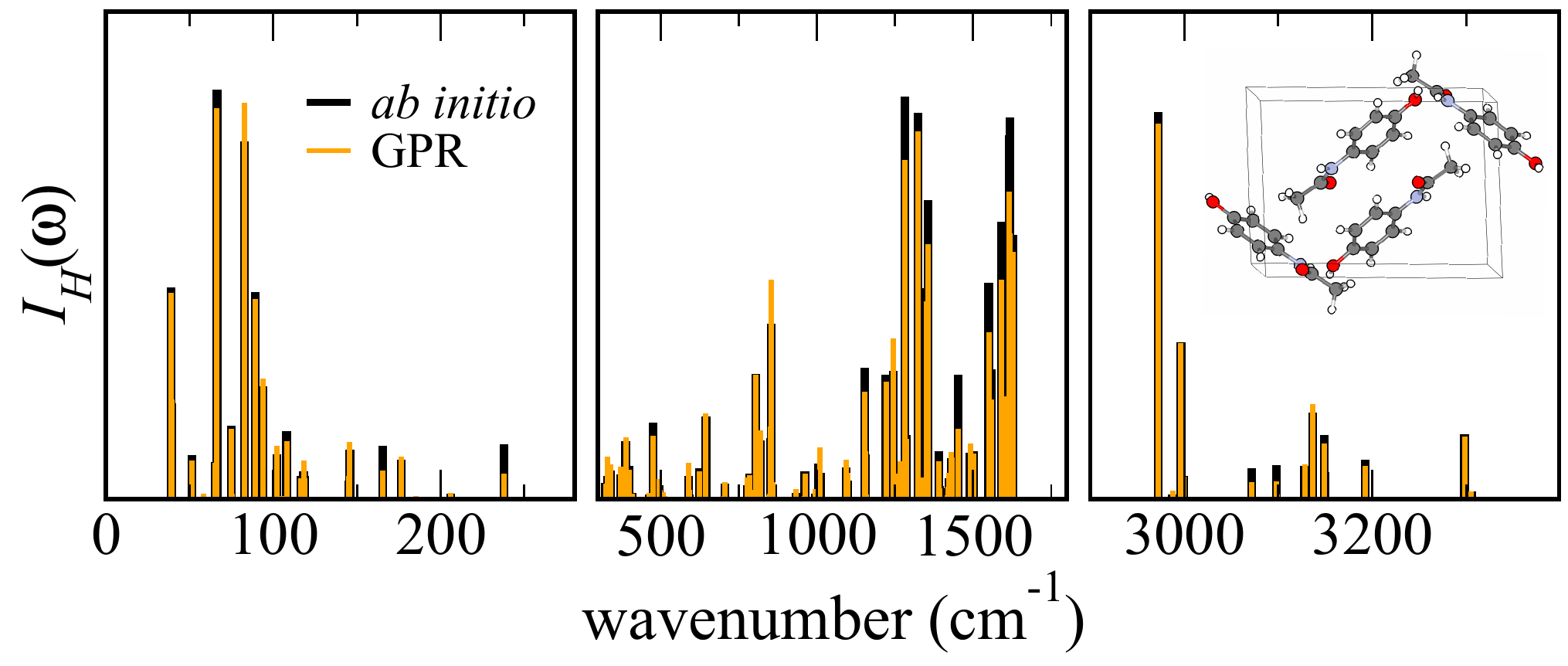}
  \caption{Harmonic Raman spectrum of Paracetamol I, obtained either via DFPT or GPR (using molecular baselining), when trained on 1500 configurations coming from an NVT trajectory. The prediction was here made on 480 displacements around the equilibrium geometry.}
  \label{fig_crystal_harmonic}
\end{figure}

We show the results for Paracetamol I in Fig.~\ref{fig_crystal_harmonic}, where we train on an NVT trajectory, as described in the main text, and use Eq.~15.
We see that the predicted harmonic spectrum reproduces nicely the \textit{ab initio} one. We note that the predicted polarizabilities experience a small global offset, which does not impact the spectrum.

Beyond this agreement, these results also signify that the computational cost of a harmonic Raman spectrum can be basically reduced to the cost of a simple geometry optimization.


\section{Offset in predictions of Paracetamol II}

\begin{figure}[hbtp]
  \centering
  \includegraphics[width=1.0\columnwidth]{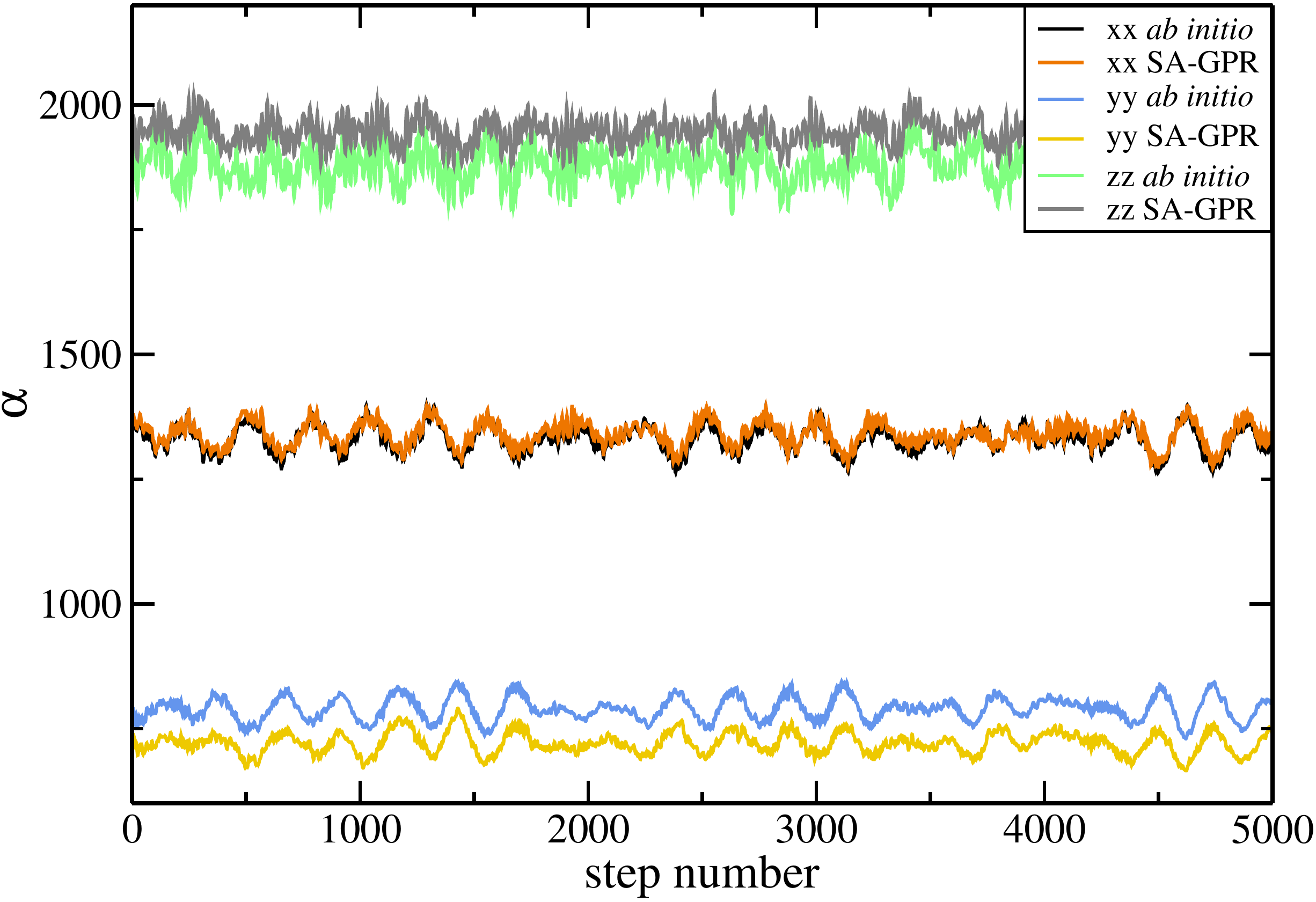}
  \caption{Polarizability time series for 3 components, obtained either with DFPT or with SA-GPR.}
  \label{fig_paraII_offset}
\end{figure}

In this section we show in Fig.~\ref{fig_paraII_offset} that the prediction of the individual susceptibility components that enter the Paracetamol II Raman spectrum presented in the main text (based on the GPR trained for form I) show different offsets with respect to the DFPT values..

\bibliography{main,refs} 
\bibliographystyle{iopart-num}